\documentclass[11pt,a4paper]{article}
\usepackage[hyperref]{acl2021}
\usepackage{times}
\usepackage{latexsym}
\usepackage{graphicx}
\usepackage{caption}
\usepackage{amsthm}
\usepackage{amssymb}
\usepackage{amsmath}
\usepackage{booktabs}
\usepackage{romannum}
\usepackage{subfigure}
\usepackage{blkarray}
\usepackage{multirow}
\usepackage{arydshln}
\usepackage{algorithm}
\usepackage{hyperref}
\usepackage{algpseudocode}

\def \Ours{SNUH}
\def\VR{\kern-\arraycolsep\strut\vrule &\kern-\arraycolsep}
\def\vr{\kern-\arraycolsep & \kern-\arraycolsep}

\usepackage{microtype}

\aclfinalcopy % Uncomment this line for the final submission
\title{Integrating Semantics and Neighborhood Information with Graph-Driven Generative Models for Document Retrieval}
\author{
	Zijing Ou$^{1}$,~ Qinliang Su$^{1}$\thanks{~~Corresponding author. Qinliang Su is also affiliated with (\romannum{1}) Guangdong Key Lab. of Big Data Analysis and Processing, Guangzhou, China, and (\romannum{2}) Key Lab. of Machine Intelligence and Advanced Computing, Ministry of Education, China.},~ Jianxing Yu$^{2}$,~ Bang Liu$^3$, \\
	\bf ~Jingwen Wang$^{4}$, ~Ruihui Zhao$^{4}$, ~Changyou Chen$^{5}$ \and Yefeng Zheng$^{4}$\\
	 $^1$School of Computer Science and Engineering, Sun Yat-sen University, Guangzhou, China, \\ 
	 $^2$School of Artificial Intelligence, Sun Yat-sen University, Guangdong, China, \\
	$^3$RALI \& Mila, Université de Montréal,  $^4$Tencent Jarvis Lab, \\
	$^5$CSE Department, SUNY at Buffalo \\
	ouzj@mail2.sysu.edu.cn, \{suqliang, yujx26\}@mail.sysu.edu.cn,  bang.liu@umontreal.ca, \\ 
	\{v\_wjwowang, zacharyzhao, yefengzheng\}@tencent.com, changyou@buffalo.edu \\
}

\date{}

\begin{document}
\maketitle
\begin{abstract}
With the need of fast retrieval speed and small memory footprint, document hashing has been playing a crucial role in large-scale information retrieval. To generate high-quality hashing code, both semantics and neighborhood information are crucial. However, most existing methods leverage only one of them or simply combine them via some intuitive criteria, lacking a theoretical principle to guide the integration process. In this paper, we encode the neighborhood information with a graph-induced Gaussian distribution, and propose to integrate the two types of information with a graph-driven generative model. To deal with the complicated correlations among documents, we further propose a tree-structured approximation method for learning. Under the approximation, we prove that the training objective can be decomposed into terms involving only singleton or pairwise documents, enabling the model to be trained as efficiently as uncorrelated ones. Extensive experimental results on three benchmark datasets show that our method achieves superior performance over state-of-the-art methods, demonstrating the effectiveness of the proposed model for simultaneously preserving semantic and neighborhood information.\footnote{Our code is available at \href{https://github.com/J-zin/SNUH}{https://github.com/J-zin/SNUH}. The MindSpore code will also be released soon.}
\end{abstract}

\section{Introduction}
Similarity search plays a pivotal role in a variety of tasks, such as image retrieval \cite{jing2008visualrank,zhang2018visual}, plagiarism detection \cite{stein2007strategies} and recommendation systems \cite{koren2008factorization}. If the search is carried out in the original continuous feature space directly, the requirements of computation and storage would be extremely high, especially for large-scale applications. Semantic hashing \cite{salakhutdinov2009semantic} sidesteps this problem by learning a compact binary code for every item such that similar items can be efficiently found according to the Hamming distance of binary codes.

Unsupervised semantic hashing aims to learn for each item a binary code that can preserve the semantic similarity information of original items, without the supervision of any labels. Motivated by the success of deep generative models \cite{salakhutdinov2009deep,kingma2013auto,rezende2014stochastic} in unsupervised representation learning, many recent methods approach this problem from the perspective of deep generative models, leading to state-of-the-art performance on benchmark datasets. Specifically, these methods train a deep generative model to model the underlying documents and then use the trained generative model to extract continuous or binary representations from the original documents \cite{chaidaroon2017variational,shen2018nash,dong2019document,zheng2020generative}. The basic principle behind these generative hashing methods is to have the hash codes retaining as much semantics information of original documents as possible so that semantically similar documents are more likely to yield similar codes.

In addition to semantics information, it is widely observed that neighborhood information among the documents is also useful to generate high-quality hash codes. By constructing an adjacency matrix from the raw features of documents, neighbor-based methods seek to preserve the information in the constructed adjacency matrix, such as the locality-preserving hashing \cite{he2004locality,zhao2014locality}, spectral hashing \cite{weiss2009spectral,li2012spectral}, and etc. However, since the ground-truth neighborhood information is not available and the constructed one is neither accurate nor complete, neighbor-based methods alone do not perform as well as the semantics-based ones. Despite both semantics and neighborhood information are derived from the original documents, different aspects are emphasized in them. Thus, to obtain higher-quality hash codes, it has been proposed to incorporate the constructed neighborhood information into semantics-based methods. For examples, \citet{chaidaroon2018deep} and \citet{hansen2020unsupervised} require the hash codes can reconstruct neighboring documents, in addition to the original input. Other works \cite{shen2019unsupervised,hansen2019unsupervised} use an extra loss term, derived from the approximate neighborhood information, to encourage similar documents to produce similar codes. However, all of the aforementioned methods exploit the neighborhood information by using it to design different kinds of regularizers to the original semantics-based models, lacking a basic principle to unify and leverage them under one framework.

To fully exploit the two types of information, in this paper, we propose a hashing method that unifies the semantics and neighborhood information with the {\emph{graph-driven generative models}}. Specifically, we first encode the neighborhood information with a multivariate Gaussian distribution. With this Gaussian distribution as a prior in a generative model, the neighborhood information can be naturally incorporated into the semantics-based hashing model. Despite the simplicity of the modeling, the correlation introduced by the neighbor-encoded prior poses a significant challenge to the training since it invalidates the widely used identical-and-independent-distributed ({\it i.i.d.}) assumption, making all documents correlated. To address this issue, we propose to use a tree-structured distribution to capture as much as possible the neighborhood information. We prove that under the tree approximation, the evidence lower bound (ELBO) can be decomposed into terms involving only singleton and pairwise documents, enabling the model to be trained as efficiently as the models without considering the document correlations. To capture more neighborhood information, a more accurate approximation by using multiple trees is also developed. Extensive experimental results on three public datasets demonstrate that the proposed method can outperform state-of-the-art methods, indicating the effectiveness of the proposed framework in unifying the semantic and neighborhood information for  document hashing.

\section{Preliminaries} \label{section:preliminaries}

\paragraph{Semantics-Based Hashing} Due to the similarities among the underlying ideas of these methods, we take the variational deep semantic hashing (VDSH) \cite{chaidaroon2017variational} as an example to illustrate their working flow. Given a document ${\boldsymbol{x}} \triangleq \{w_j\}_{j=1}^{|{\boldsymbol{x}}|}$,  VDSH proposes to model a document by a generative model as
\begin{align}
	p({\boldsymbol{x}}, {\boldsymbol{z}}) =  p_{\boldsymbol{\theta}}({\boldsymbol{x}}|{\boldsymbol{z}}) p({\boldsymbol{z}}),
\end{align}
where $p({\boldsymbol{z}})$ is the prior distribution and is chosen to be the standard Gaussian distribution ${\mathcal{N}}({\boldsymbol{z}}; {\boldsymbol{0}}, {\boldsymbol{I}}_d)$, with ${\boldsymbol{I}}_d$ denoting the $d$-dimensional identity matrix; and $p_{\boldsymbol{\theta}}({\boldsymbol{x}}|{\boldsymbol{z}})$ is defined to be
\begin{align}\label{likelihood_func_x}
	p_{\boldsymbol{\theta}}({\boldsymbol{x}}|{\boldsymbol{z}}) = \prod\limits_{w_i \in \boldsymbol{x}} p_{\boldsymbol{\theta}}(w_i | \boldsymbol{z})
\end{align}
with
\begin{equation}
	\label{decoder}
	p_{\boldsymbol{\theta}}(w_i | \boldsymbol{z}) \triangleq \frac{\text{exp}(\boldsymbol{z}^T E w_i + b_i)}{\sum_{j  = 1}^{|V|} \text{exp}(\boldsymbol{z}^T E w_j + b_j)},
\end{equation}
in which $w_j$ denotes the $|V|$-dimensional one-hot representation of the $j$-th word, with $|{\boldsymbol{x}}|$ and $|V|$ denoting the document and vocabulary size, respectively; and $E \in \mathbb{R}^{d \times |V|}$ represents the learnable embedding matrix. For a corpus containing $N$ documents ${\boldsymbol{X}} = \{{\boldsymbol{x_1}}, {\boldsymbol{x}}_2, \cdots, {\boldsymbol{x}}_N\}$, due to the {\it i.i.d.} assumption for documents, it is modelled by simply multiplying individual document models as
\begin{equation}
	p({\boldsymbol{X}}, {\boldsymbol{Z}}) = \prod_{k=1}^N p_{\boldsymbol{\theta}}({\boldsymbol{x}}_k|{\boldsymbol{z}}_k)p({\boldsymbol{z}}_k),
\end{equation}
where ${\boldsymbol{Z}} \triangleq [{\boldsymbol{z}}_1; {\boldsymbol{z}}_2; \cdots; {\boldsymbol{z}}_N]$ denotes a long vector obtained by concatenating the individual vectors ${\boldsymbol{z}}_i$. The model is trained by optimizing the evidence lower bound (ELBO) of the log-likelihood function $\log p({\boldsymbol{X}})$. After training, outputs from the trained encoder are used as documents' representations, from which binary hash codes can be obtained by thresholding the real-valued representations.

\paragraph{Neighborhood Information} The ground-truth semantic similarity information is not available for the unsupervised hashing task in practice. To leverage this information, an affinity $N\times N$ matrix ${\boldsymbol{A}}$ is generally constructed from the raw features ({\it e.g.}, the TFIDF) of original documents. For instances, we can construct the matrix as
\begin{equation} \label{affinity_matrix}
	a_{i j} \!=\! \left\{\begin{array}{ll}
		e^{-\frac{\left|\left|{\boldsymbol{x}}_i - {\boldsymbol{x}}_j\right|\right|^2}{\sigma}}, & \boldsymbol{x}_{i} \! \in \! \mathcal{N}_{k}\left(\boldsymbol{x}_{j}\right)  \\
		0, & \text { otherwise }
	\end{array}\right.
\end{equation}
where $a_{ij}$ denotes the $(i, j)$-th element of ${\boldsymbol{A}}$; and $\mathcal{N}_{k}(\boldsymbol{x})$ denotes the $k$-nearest neighbors of document $\boldsymbol{x}$. Given the affinity matrix ${\boldsymbol{A}}$, some methods have been proposed to incorporate the neighborhood information into the semantics-based hashing models. However, as discussed above, these methods generally leverage the information based on some intuitive criteria, lacking theoretical supports behind them.

\section{A Hashing Framework with Unified Semantics-Neighborhood Information}

In this section, we present a more effective framework to unify the semantic and neighborhood information for the task of document hashing.

\subsection{Reformulating the VDSH}
To introduce the neighborhood information into the semantics-based hashing models, we first rewrite the VDSH model into a compact form as
\begin{align} \label{VDSH_model}
	p({\boldsymbol{X}}, {\boldsymbol{Z}}) = p_{\boldsymbol{\theta}}({\boldsymbol{X}}| {\boldsymbol{Z}}) p_{\mathcal{I}}({\boldsymbol{Z}}),
\end{align}
where $p_{\boldsymbol{\theta}}({\boldsymbol{X}}| {\boldsymbol{Z}})=\prod_{k=1}^N p_{\boldsymbol{\theta}}({\boldsymbol{x}}_k|{\boldsymbol{z}}_k)$; and the prior $p_{\mathcal{I}}({\boldsymbol{Z}}) = \prod_{k=1}^N p({\boldsymbol{z}}_k)$, which can be shown to be
\begin{equation}
	p_{\mathcal{I}}({\boldsymbol{Z}}) = {\mathcal{N}}\left({\boldsymbol{Z}}; {\boldsymbol{0}}, {\boldsymbol{I}}_N  \otimes {\boldsymbol{I}}_d \right).
\end{equation}
Here, $\otimes$ denotes the Kronecker product and the subscript ${\mathcal{I}}$ indicates independence among ${\boldsymbol{z}}_k$. The ELBO of this model can be expressed as
\begin{align}
	{\mathcal{L}} \!\!= \! \underbrace {{\mathbb{E}}_{q_{\boldsymbol{\phi}}( \! {\boldsymbol{Z}}|{\boldsymbol{X}}\!)} \!\! \left[ \log p_{\boldsymbol{\theta}} (\!{\boldsymbol{X}}|{\boldsymbol{Z}})  \right]}_{{\mathcal{L}}_1}
	\!\!- \!\! \underbrace{KL\! \left(\! q_{\boldsymbol{\phi}}(\!{\boldsymbol{Z}}|{\boldsymbol{X}}\!) || p_{\mathcal{I}}(\!{\boldsymbol{Z}}) \right)}_{{\mathcal{L}}_2} \nonumber
\end{align}
where $KL(\cdot)$ denotes the Kullback-Leibler (KL) divergence. By restricting the posterior to independent Gaussian form
\begin{align}
	q_{\boldsymbol{\phi}}({\boldsymbol{Z}}|{\boldsymbol{X}}) &= \prod_{k=1}^N \underbrace {{\mathcal{N}}\left({\boldsymbol{z}}_k; {\boldsymbol{\mu}}_k, diag({\boldsymbol{\sigma}}^2_k)\right)}_{q_{\boldsymbol{\phi}}({\boldsymbol{z}}_k|{\boldsymbol{x}}_k)}, \label{q_phi_ZX}
\end{align}
the ${\mathcal{L}}_1$ can be handled using the reparameterization trick. Thanks to the factorized forms assumed in $q_{\boldsymbol{\phi}}({\boldsymbol{Z}}|{\boldsymbol{X}})$ and  $p_{\mathcal{I}}({\boldsymbol{Z}})$, the ${\mathcal{L}}_2$ term can also be expressed analytically and evaluated efficiently.

\subsection{Injecting the Neighborhood Information}
Given an affinity matrix ${\boldsymbol{A}}$, the covariance matrix ${\boldsymbol{I}}_N + \lambda {\boldsymbol{A}}$ can be used to reveal the neighborhood information of documents, where the hyperparameter $\lambda \in [0, 1)$ is used to control the overall correlation strength. If two documents are neighboring, then the corresponding correlation value in ${\boldsymbol{I}}_N + \lambda {\boldsymbol{A}}$ will be large; otherwise, the value will be zero. To have the neighborhood information reflected in  document representations, we can require that the representations ${\boldsymbol{z}}_i$ are drawn from a Gaussian distribution of the form
\begin{align} \label{prior_Z}
	p_{{\mathcal{G}}}(\boldsymbol{Z}) = {\mathcal{N}}\left({\boldsymbol{Z}}; {\boldsymbol{0}}, \left( {\boldsymbol{I}}_N + \lambda {\boldsymbol{A}} \right)
	\otimes {\boldsymbol{I}}_d \right),
\end{align}
where the subscript ${\mathcal{G}}$ denotes that the distribution is constructed from a neighborhood graph. To see why the representations ${\boldsymbol{Z}} \sim p_{\mathcal{G}}({\boldsymbol{Z}})$ have already reflected the neighborhood information, let us consider an example with three documents $\{{\boldsymbol{x}}_1, {\boldsymbol{x}}_2, {\boldsymbol{x}}_3\}$, in which ${\boldsymbol{x}}_1$ is connected to ${\boldsymbol{x}}_2$, ${\boldsymbol{x}}_2$ is connected to ${\boldsymbol{x}}_3$, and no connection exists between ${\boldsymbol{x}}_1$ and ${\boldsymbol{x}}_3$. Under the case that ${\boldsymbol{z}}_i$ is a two-dimensional vector ${\boldsymbol{z}}_i \in {\mathbb{R}}^2$, we have the concatenated representations $[{\boldsymbol{z}}_1; {\boldsymbol{z}}_2; {\boldsymbol{z}}_3]$ follow a Gaussian distribution with covariance matrix of
% \begin{equation*}
% \begin{array}{ccc:cc:ccc}
% & \multicolumn{2}{c}{\boldsymbol{z}_1} & \multicolumn{2}{c}{\boldsymbol{z}_2} & \multicolumn{2}{c}{\boldsymbol{z}_2} & \\
% \multirow{2}{*}{$\boldsymbol{z}_1$}
% & 1 & 0 & {\lambda a_{12}} & 0 & 0 & 0 & \\
% & 0 & 1 & 0 & {\lambda a_{12}} & 0 & 0 & \\ \cdashline{2-7}
% \multirow{2}{*}{$\boldsymbol{z}_2$}
% & {\lambda a_{21}} & 0 & 1 & 0 & {\lambda a_{23}} & 0 & \\
% & 0 & {\lambda a_{21}} & 0 & 1 & 0 & {\lambda a_{23}} & \\ \cdashline{2-7}
% \multirow{2}{*}{$\boldsymbol{z}_3$}
% & 0 & 0 & {\lambda a_{32}} & 0 & 1 & 0 & \\
% & 0 & 0 & 0 & {\lambda a_{32}} & 0 & 1 & .
% \end{array}
% \end{equation*}
\resizebox{.48\textwidth}{!}{ 
\begin{blockarray}{ccccccc}
& \BAmulticolumn{2}{c}{$\boldsymbol{z}_1$} & \BAmulticolumn{2}{c}{$\boldsymbol{z}_2$} & \BAmulticolumn{2}{c}{$\boldsymbol{z}_3$} \\
\begin{block}{c(cc|cc|cc)}
\multirow{2}{*}{$\boldsymbol{z}_1$}
& 1 & 0 & {$\lambda a_{12}$} & 0 & 0 & 0 \\
& 0 & 1 & 0 & {$\lambda a_{12}$} & 0 & 0 \\ \cline{2-7}
\multirow{2}{*}{$\boldsymbol{z}_2$}
& {$\lambda a_{21}$} & 0 & 1 & 0 & {$\lambda a_{23}$} & 0 \\
& 0 & {$\lambda a_{21}$} & 0 & 1 & 0 & {$\lambda a_{23}$} \\ \cline{2-7}
\multirow{2}{*}{$\boldsymbol{z}_3$}
& 0 & 0 & {$\lambda a_{32}$} & 0 & 1 & 0 \\
& 0 & 0 & 0 & {$\lambda a_{32}$} & 0 & 1 \\
\end{block}
\end{blockarray} }
% \begin{equation*}
% \small
% \bordermatrix{
%   & \boldsymbol{z}_1 & \vr \boldsymbol{z}_2 & \vr \boldsymbol{z}_3 \cr
%  \boldsymbol{z}_1 & \begin{array}{cc}
%      1 & 0 \\
%      0 & 1
%  \end{array} & \VR \begin{array}{cc}
%      {\lambda a_{12}} & 0 \\
%      0 & {\lambda a_{12}}
%  \end{array} & \VR \begin{array}{cc}
%      0 & 0 \\
%      0 & 0
%  \end{array} \cr \cline{2-6}
%  \boldsymbol{z}_2 & \begin{array}{cc}
%      {\lambda a_{21}} & 0 \\
%      0 & {\lambda a_{21}}
%  \end{array} & \VR \begin{array}{cc}
%      1 & 0 \\
%      0 & 1
%  \end{array} & \VR \begin{array}{cc}
%      {\lambda a_{23}} & 0 \\
%      0 & {\lambda a_{23}}
%  \end{array} \cr \cline{2-6}
%  \boldsymbol{z}_3 & \begin{array}{cc}
%      0 & 0 \\
%      0 & 0
%  \end{array} & \VR \begin{array}{cc}
%      {\lambda a_{32}} & 0 \\
%      0 & {\lambda a_{32}}
%  \end{array} & \VR \begin{array}{cc}
%      1 & 0 \\
%      0 & 1
%  \end{array} 
% }.
% \end{equation*}
From the property of Gaussian distribution, it can be known that ${\boldsymbol{z}}_1$ is strongly correlated  with  ${\boldsymbol{z}}_2$ on the corresponding elements, but not with ${\boldsymbol{z}}_3$. This suggests that ${\boldsymbol{z}}_1$ should be similar to ${\boldsymbol{z}}_2$, but different from ${\boldsymbol{z}}_3$, which is consistent with the neighborhood relation that ${\boldsymbol{x}}_1$ is a neighbor of ${\boldsymbol{x}}_2$, but not of ${\boldsymbol{x}}_3$.

Now that the neighborhood information can be modeled by requiring ${\boldsymbol{Z}}$ being drawn from $p_{\mathcal{G}}({\boldsymbol{Z}})$, and the semantic information can be reflected in the likelihood function $p_{\boldsymbol{\theta}}({\boldsymbol{X}}| {\boldsymbol{Z}})$. The two types of information can be taken into account simultaneously by modeling the corpus as
\begin{equation}
	p({\boldsymbol{X}}, {\boldsymbol{Z}}) = p_{\boldsymbol{\theta}}({\boldsymbol{X}}| {\boldsymbol{Z}}) p_{{\mathcal{G}}}(\boldsymbol{Z}).
\end{equation}
Comparing to the VDSH model in \eqref{VDSH_model}, it can be seen that the only difference lies in the employed priors. Here, a neighborhood-preserving prior $p_{\mathcal{G}}({\boldsymbol{Z}})$ is employed, while in VDSH, an independent prior $p_{\mathcal{I}}({\boldsymbol{Z}})$ is used. Although only a modification to the prior is made from the perspective of modeling, significant challenges are posed for the training. Specifically, by replacing $p_{\mathcal{I}}({\boldsymbol{Z}})$ with $p_{\mathcal{G}}({\boldsymbol{Z}})$ in the ${\mathcal{L}}_2$ of $\mathcal{L}$, it can be shown that the expression of ${\mathcal{L}}_2$ involves the matrix $\big(\left( {\boldsymbol{I}}_N \!\! + \!\! \lambda {\boldsymbol{A}} \right) 
\! \otimes \! {\boldsymbol{I}}_d\big)^{-1}$. Due to the introduced dependence  among documents, for example, if the corpus contains over 100,000 documents and the representation dimension is set to 100, the ${\mathcal{L}}_2$ involves the {\em inverse} of matrices with dimension as high as $10^7$, which is computationally prohibitive in practice.

\section{Training with Tree Approximations}

Although the prior $p_{\mathcal{G}}({\boldsymbol{Z}})$ captures the full neighborhood information, its induced model is not practically trainable. In this section, to facilitate the training, we first propose to use a tree-structured prior to partially capture the neighborhood information, and then extend it to multiple-tree case for more accurate modeling.

\subsection{Approximating the Prior $p_{\mathcal{G}}({\boldsymbol{Z}})$ with a Tree-Structured Distribution}

The matrix ${\boldsymbol{A}}$ represents a graph ${\mathbb{G}} \triangleq ({\mathcal{V}}, {\mathcal{E}})$, where ${\mathcal{V}} = \{1, 2, \cdots, N\}$ is the set of document indices; and ${\mathcal{E}} = \{(i, j)| a_{ij} \ne 0 \}$ is the set of connections between documents. From the graph ${\mathbb{G}}$, a spanning tree ${\mathbb{T}} = ({\mathcal{V}}, {\mathcal{E}}_T)$ can be obtained easily, where ${\mathcal{E}}_T$ denotes the set of connections on the tree.\footnote{We assume the graph is connected. For more general cases, results can be derived similarly.} Based on the spanning tree, we construct a new distribution as
\begin{equation} \label{p_T(z)}
	\!p_{\mathcal{T}}(\boldsymbol{Z}) = \prod_{i\in {\mathcal{V}}}  p_{\mathcal{G}}(\boldsymbol{z}_i) \!\! \prod_{(i, j)\in {\mathcal{E}}_T} \frac{p_{\mathcal{G}}(\boldsymbol{z}_i , \boldsymbol{z}_j)}{p_{\mathcal{G}}(\boldsymbol{z}_i) p_{\mathcal{G}}(\boldsymbol{z}_j)},
\end{equation}
where $p_{\mathcal{G}}({\boldsymbol{z}}_i)$ and $p_{\mathcal{G}}({\boldsymbol{z}}_i, {\boldsymbol{z}}_j)$ represent one- and two-variable marginal distributions of $p_{\mathcal{G}}({\boldsymbol{Z}})$, respectively. From the properties of Gaussian distribution, it is known that
\begin{align} \label{correlated_prior}
	\!\!\!p_{\mathcal{G}}({\boldsymbol{z}}_i) \! &= \! {\mathcal{N}} \! ({\boldsymbol{z}}_i; {\boldsymbol{0}}, {\boldsymbol{I}}_d ), \nonumber \\
	\!\!\!p_{\mathcal{G}}({\boldsymbol{z}}_i, {\boldsymbol{z}}_j) \! &= \! {\mathcal{N}} \!\! \left(  [{\boldsymbol{z}}_i; \! {\boldsymbol{z}}_j]; {\boldsymbol{0}}, \! ({\boldsymbol{I}}_2 \! + \! \lambda {\boldsymbol{A}_{ij}}) \! \otimes \! {\boldsymbol{I}}_d \! \right),
\end{align}
where ${\boldsymbol{A}}_{ij} \triangleq \left[ {\begin{array}{*{5}{c}}
 \!\!\! 0 & \!\!\! a_{ij} \!\!\! \\ 
\!\!\!  a_{ji} & \!\!\! 0 \!\!\!
\end{array}} \right]$. Because $p_{\mathcal{T}}(\boldsymbol{Z})$ is defined on a tree, as proved in \cite{wainwright2008graphical}, it is guaranteed to be a valid probability distribution, and more importantly, it satisfies the following two relations: \romannum{1}) $p_{\mathcal{T}}(\boldsymbol{z}_i) = p_{\mathcal{G}}(\boldsymbol{z}_i)$; \romannum{2}) $p_{\mathcal{T}}(\boldsymbol{z}_i, \boldsymbol{z}_j) = p_{\mathcal{G}}(\boldsymbol{z}_i, \boldsymbol{z}_j)$ for any $(i, j) \in {\mathcal{E}}_T$, where $p_{\mathcal{T}}(\boldsymbol{z}_i)$ and $p_{\mathcal{T}}(\boldsymbol{z}_i, \boldsymbol{z}_j)$ denote the marginal distributions of $p_{\mathcal{T}}(\boldsymbol{Z})$. That is, the tree-structured distribution $p_{\mathcal{T}}(\boldsymbol{Z})$ captures the neighborhood information reflected on the spanning tree ${\mathbb{T}}$. By using $p_{\mathcal{T}}({\boldsymbol{Z}})$ to replace $p_{\mathcal{I}}({\boldsymbol{Z}})$ of ${\mathcal{L}}_2$, it can be shown that ${\mathcal{L}}_2$ can be expressed as the summation of terms involving only one or two variables, which can be handled easily. Due to the limitation of space, the concrete expression for the lower bound is given in the Supplementary Material.

\subsection{Imposing Correlations on the Posterior}
The posterior distribution $q_{\boldsymbol{\phi}}({\boldsymbol{Z}}| {\boldsymbol{X}})$ in the previous section is assumed to be in independent form, as the form shown in \eqref{q_phi_ZX}. But since a prior $p_{\mathcal{T}}({\boldsymbol{Z}})$ considering the correlations among documents is used, assuming an independent posterior is not appropriate. Hence, we follow the tree-structured prior and also construct a tree-structured posterior
\begin{equation} \label{q_T(Z)}
     q_{\mathcal{T}}(\boldsymbol{Z} | \boldsymbol{X}) 
    \!\!= \!\! \prod_{i\in {\mathcal{V}}} \!\!q_{\boldsymbol{\phi}}\! \left(\boldsymbol{z}_{i} | \boldsymbol{x}_{i}\right) \!\!\!\!\!\! \prod_{(i, j)\in {\mathcal{E}}_T} \!\!\!\! \frac{q_{\boldsymbol{\phi}}\left(\boldsymbol{z}_{i}, \boldsymbol{z}_{j} | \boldsymbol{x}_{i}, \boldsymbol{x}_{j}\right)}{q_{\boldsymbol{\phi}}\! \left(\boldsymbol{z}_{i} | \boldsymbol{x}_{i}\right) \! q_{\boldsymbol{\phi}}\!\left(\boldsymbol{z}_{j} | \boldsymbol{x}_{j}\right)}, \nonumber
\end{equation}
where $q_{\boldsymbol{\phi}}({\boldsymbol{z}}_i|{\boldsymbol{x}}_i)$ is the same as that in \eqref{q_phi_ZX}; and $q_{\boldsymbol{\phi}}\left(\boldsymbol{z}_i, \boldsymbol{z}_j | \boldsymbol{x}_i, \boldsymbol{x}_j\right)$ is also defined to be Gaussian, with its mean defined as $[{\boldsymbol{\mu}}_i; {\boldsymbol{\mu}}_j]$ and covariance matrix defined as
\begin{align} \label{q_zi_zi}
\left[\!\!\! \begin{array}{cc}
	diag(\boldsymbol{\sigma}_{i}^2) \!\!\! & \!\!\! diag(\boldsymbol{\gamma}_{ij} \! \odot \! \boldsymbol{\sigma}_{i} \! \odot \! \boldsymbol{\sigma}_{j}) \\
	diag(\boldsymbol{\gamma}_{ij} \! \odot \! \boldsymbol{\sigma}_{i} \! \odot \! \boldsymbol{\sigma}_{j}) \!\!\! & \!\!\! diag(\boldsymbol{\sigma}_{j}^2)
\end{array} \!\!\! \right],
\end{align}
   in which $\boldsymbol{\gamma}_{ij} \in \mathbb{R}^{d}$ controls the correlation strength between $\boldsymbol{z}_i$ and $\boldsymbol{z}_j$, whose elements are restricted in $(-1, 1)$ and $\odot$ denotes the Hadamard product. By taking the correlated posterior $q_{\mathcal{T}}(\boldsymbol{Z}|\boldsymbol{X})$ into the ELBO, we obtain
\begin{equation}
\begin{aligned}
    \mathcal{L}_\mathcal{T} &\!= \!\!  \sum_{i\in {\mathcal{V}}}  \!{\mathbb{E}}_{q_{\boldsymbol{\phi}}} \!\! \left[ \log p_{\boldsymbol{\theta}} ({\boldsymbol{x}_i}|{\boldsymbol{z}_i}) \right] \!\!- \!\! K\!L \! \left( q_{\boldsymbol{\phi}}(\boldsymbol{z}_i) || p_{\mathcal{G}}(\boldsymbol{z}_i) \right) \nonumber \\
    &\!- \!\!\!\!\!\! \sum_{(i, j)\in {\mathcal{E}}_T} \!\!\!\! \Big( K\!L\left( q_{\boldsymbol{\phi}}(\boldsymbol{z}_i , \boldsymbol{z}_j | \boldsymbol{x}_i, \boldsymbol{x}_j) || p_{\mathcal{G}}(\boldsymbol{z}_i, \boldsymbol{z}_j) \right) \nonumber \\
    &\!- \!\! K\!L\! \left( q_{\boldsymbol{\phi}}(\boldsymbol{z}_i) || p_{\mathcal{G}}(\boldsymbol{z}_i) \right)  \!\!- \!\! K\!L\! \left( q_{\boldsymbol{\phi}}(\boldsymbol{z}_j) || p_{\mathcal{G}}(\boldsymbol{z}_j) \right) \!\! \Big), \nonumber
\end{aligned}
\end{equation}
where we briefly denote the variational distribution $q_{\boldsymbol{\phi}}(\boldsymbol{z}_i | \boldsymbol{x}_i)$ as $q_{\boldsymbol{\phi}}(\boldsymbol{z}_i)$. 
Since $p_{\mathcal{G}}({\boldsymbol{z}}_i)$,  $p_{\mathcal{G}}({\boldsymbol{z}}_i, {\boldsymbol{z}}_j)$, $q_{\boldsymbol{\phi}}(\boldsymbol{z}_i | \boldsymbol{x}_i)$ and $q_{\boldsymbol{\phi}}({\boldsymbol{z}}_i, {\boldsymbol{z}}_j|{\boldsymbol{x}}_i, {\boldsymbol{x}}_j)$ are all Gaussian distributions,  the KL-divergence terms above can be derived in closed-form. Moreover, it can be seen that ${\mathcal{L}}_{\mathcal{T}}$ involves only single or pairwise variables, thus optimizing it is as efficient as the models without considering document correlation.

With the trained model, hash codes can be obtained by binarizing the posterior mean ${\boldsymbol{\mu}}_i$ with a threshold, as done in \cite{chaidaroon2017variational}. However, if without any constraint, the range of mean lies in $(-\infty, +\infty)$. Thus, if we binarize it directly, lots of information in the original representations will be lost. To alleviate this problem, in our implementation, we parameterize the posterior mean ${\boldsymbol{\mu}}_i$ by a function of the form ${\boldsymbol{\mu}}_i =sigmoid (nn(\boldsymbol{x}_i )/ \tau)$, where the outermost sigmoid function forces the mean to look like binary value and thus can effectively reduce the quantization loss, with $nn(\cdot)$ denoting a neural network function and $\tau$ controlling the slope of the sigmoid function.

\subsection{Extending to Multiple Spanning Trees}
Obviously, approximating the graph with a spanning tree may lose too much information. To alleviate this issue, we propose to capture the similarity information by a mixture of multiple distributions, with each built on a spanning tree. Specifically, we first construct a set of $M$ spanning trees ${\mathcal{T}}_{\mathbb{G}} = \{{\mathbb{T}}_1, {\mathbb{T}}_2, \cdots, {\mathbb{T}}_M \} $ from the original graph ${\mathbb{G}}$. Based on the set of spanning trees, a mixture-distribution prior and posterior can be constructed as
\begin{align}
    p_{\mathcal{MT}}(\boldsymbol{Z}) &= \frac{1}{M} \sum\limits_{{\mathcal{T}} \in \mathcal{T}_{\mathbb{G}}} p_{\mathcal{T}}(\boldsymbol{Z}), \\
    q_{\mathcal{MT}}(\boldsymbol{Z} | \boldsymbol{X}) &= \frac{1}{M} \sum\limits_{{\mathcal{T}} \in \mathcal{T}_{\mathbb{G}}} q_{\mathcal{T}}(\boldsymbol{Z} | \boldsymbol{X}),
\end{align}
where $p_{\mathcal{T}}({\boldsymbol{Z}})$ and $q_{\mathcal{T}}(\boldsymbol{Z} | \boldsymbol{X})$ are the prior and posterior defined on the tree ${\mathcal{T}}$, as done in \eqref{p_T(z)} and \eqref{q_T(Z)}.
By taking the mixture distributions above into the ELBO of $\mathcal{L}$ to replace the prior and posterior, we can obtain a new ELBO, denoted as ${\mathcal{L}}_{{\mathcal{M T}}}$. Obviously, it is impossible to obtain a closed-form expression for the bound ${\mathcal{L}}_{{\mathcal{M T}}}$. But as proved in \cite{tang2019correlated}, by using the log-sum inequality, ${\mathcal{L}}_{{\mathcal{M T}}}$ can be further lower bounded by
\begin{align} \label{ELBO_multi_Tree}
    \mathcal{\widetilde L }_{\mathcal{MT}} = \frac{1}{M} \sum_{{\mathcal{T}}\in {\mathcal{T}}_{\mathbb{G}}}{\mathcal{L}}_{\mathcal{T}}.
\end{align}
Given the expression of ${\mathcal{L}}_{\mathcal{T}}$, the lower bound of $\mathcal{\widetilde L }_{\mathcal{MT}} $ can also be expressed in closed-form and optimized efficiently. For detailed derivations and concrete expressions, please refer to the Supplementary.

\subsection{Details of Modeling}
The parameters $\boldsymbol{\mu}_{i}, \boldsymbol{\mu}_{j}, \boldsymbol{\sigma}_i, \boldsymbol{\sigma}_j$ and $\boldsymbol{\gamma}_{ij}$ in the approximate posterior distribution $q_{\boldsymbol{\phi}}({\boldsymbol{z}}_i| {\boldsymbol{x}}_i)$ of \eqref{q_phi_ZX} and $q_{\boldsymbol{\phi}}({\boldsymbol{z}}_i, {\boldsymbol{z}}_j| {\boldsymbol{x}}_i, {\boldsymbol{x}}_j)$ of \eqref{q_zi_zi} are all defined as the outputs of neural networks, with the parameters denoted as ${\boldsymbol{\phi}}$. Specifically, the entire model is mainly composed of three components:
\begin{itemize}
    \item[\romannum{1})] The variational encoder $q_{\boldsymbol{\phi}} (\boldsymbol{z}_i | \boldsymbol{x}_i)$, which takes single document as input, and outputs the mean and variance of Gaussian distribution, {\it i.e.,} $[\boldsymbol{\mu}_i; \boldsymbol{\sigma}_i^2] = f_{\boldsymbol{\phi}}(\boldsymbol{x}_i)$;
    \item[\romannum{2})] The correlated encoder, which takes pairwise documents as input, and outputs the correlation coefficient, {\it i.e.}, $\boldsymbol{\gamma}_{ij} = f_{\boldsymbol{\phi}}(\boldsymbol{x}_i, \boldsymbol{x}_j)$. Note that the correlation encoder is required to be order-irrelevant, that is, $f_{\boldsymbol{\phi}}(\boldsymbol{x}_i, \boldsymbol{x}_j) = f_{\boldsymbol{\phi}}(\boldsymbol{x}_j, \boldsymbol{x}_i)$, which is achieved in this paper as $f_{\boldsymbol{\phi}} = \frac{1}{2} \big( f_{\boldsymbol{\phi}}(\boldsymbol{x}_i, \boldsymbol{x}_j) + f_{\boldsymbol{\phi}}(\boldsymbol{x}_j, \boldsymbol{x}_i) \big)$;
    \item[\romannum{3})] The generative decoder $p_{\boldsymbol{\theta}}(\boldsymbol{x}_i | \boldsymbol{z}_i)$, which takes the latent variable ${\boldsymbol{z}}_i$ as input and output the document ${\boldsymbol{x}}_i$. The decoder is modeled by a neural network parameterized by ${\boldsymbol{\theta}}$.
\end{itemize}
The model is trained by optimizing the lower bound $\widetilde{ \mathcal{L}}_{\mathcal{MT}}$ w.r.t. ${\boldsymbol{\phi}}$ and ${\boldsymbol{\theta}}$. After training, hash codes are obtained by passing the documents through the variational encoder and binarizing the outputs on every dimension by a the threshold value, which is simply set as 0.5 in our experiments.

\begin{figure}[!t]
	\centering
	\includegraphics[scale=0.25]{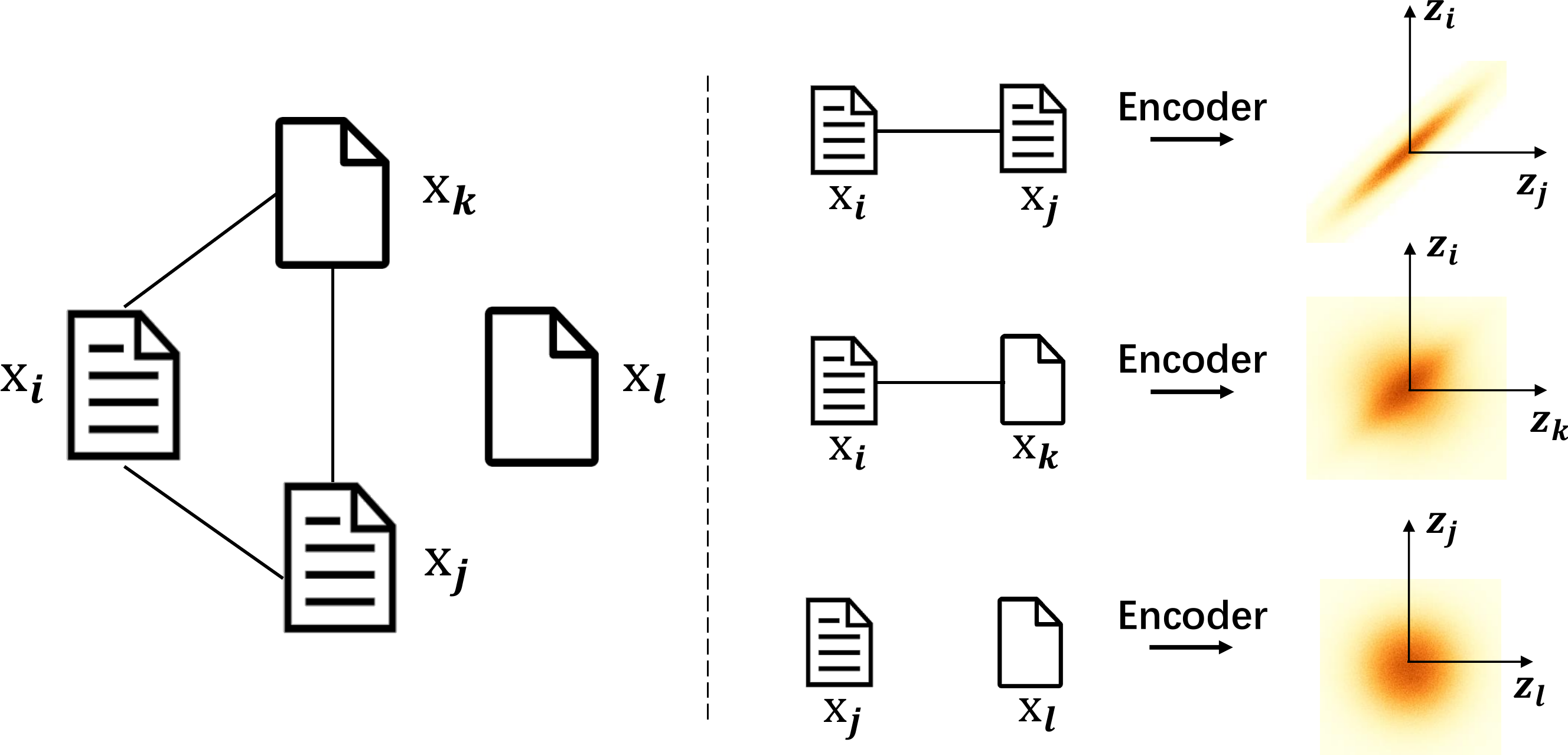}
	\vspace{-1.mm}
	\caption{Illustration of how the proposed model preserves the semantic and similarity information in the representations, where the color and link represent semantic similarity and neighborhood, respectively.}
	\vspace{-4mm}
	\label{citation_network}
\end{figure}

To intuitively understand the insight behind our model, an illustration is shown in Figure \ref{citation_network}. We see that if  the  two  documents  are  neighbors  and  semantically similar, the representations will be strongly correlated to each other.  But if they are not semantically similar neighbors, the representations become less correlated. If they are neither neighbors nor semantically similar, the representations become not correlated at all. Since our model can simultaneously preserve semantics and neighborhood information, we name it as \textbf{S}emantics-\textbf{N}eighborhood \textbf{U}nified \textbf{H}ahing (\Ours).

\section{Related Work}

Deep generative models \cite{rezende2014stochastic} have attracted a lot of attention in semantics-based hashing, due to their successes in unsupervised representation learning. VDSH \cite{chaidaroon2017variational} first employed variational auto-encoder (VAE) \cite{kingma2013auto} to learn continuous representations of documents and then casts them into binary codes. However, for the sake of information leaky problem during binarization step, such a two-stage strategy is prone to result in local optima and undermine the performance. NASH \cite{shen2018nash} tackled this issue by replacing the Gaussian prior with Bernoulli and adopted the straight-through technique \cite{bengio2013estimating} to achieve end-to-end training. To further improve the model’s capability, \citet{dong2019document} proposed to employ mixture distribution as a priori knowledge and \citet{zheng2020generative} exploited Boltzmann posterior to introduce correlation among bits. Beyond generative frameworks, AMMI \cite{stratos2020learning} achieved superior performance by maximizing the mutual information between codes and documents. Nevertheless, the aforementioned semantic hashing methods are consistently under the {\it i.i.d.} assumption, which means they ignore the neighborhood information.

Spectral hashing \cite{weiss2009spectral} and self-taught hashing \cite{zhang2010self} are two typical methods of neighbor-based hashing models. But these algorithms generally ignore the rich semantic information associated with documents.
Recently, some VAE-based models tried to concurrently take account of semantic and neighborhood information, such as NbrReg \cite{chaidaroon2018deep}, RBSH \cite{hansen2019unsupervised} and PairRec\cite{hansen2020unsupervised}. However, as mentioned before, all of them simply regarded the proximity as regularization, lacking theoretical principles to guide the incorporation process. Thanks to the virtue of graph-induced distribution, we effectively preserve the two types of information in a theoretical framework.

\section{Experiments}
\begin{table*}[!t]
	\centering
	\small
	\setlength{\tabcolsep}{1.2mm}{
		\begin{tabular}{c|cccc|cccc|cccc|c}
		\toprule
		\multirow{2}*{\textbf{Method}} & \multicolumn{4}{c}{Reuters} & \multicolumn{4}{|c}{TMC} & \multicolumn{4}{|c|}{20Newsgroups} & \multirow{2}*{Avg}  \\
		\cmidrule(r){2-5} \cmidrule(r){6-9} \cmidrule(r){10-13}
		 & 16bits & 32bits & 64bits & 128bits & 16bits & 32bits & 64bits & 128bits & 16bits & 32bits & 64bits & 128bits  \\
		\midrule
		SpH & 0.6340 & 0.6513 & 0.6290 & 0.6045 & 0.6055 & 0.6281 & 0.6143 & 0.5891 & 0.3200 & 0.3709 & 0.3196 & 0.2716 & 0.5198 \\
		STH & 0.7351 & 0.7554 & 0.7350 & 0.6986 & 0.3947 & 0.4105 & 0.4181 & 0.4123 & 0.5237 & 0.5860 & 0.5806 & 0.5443 & 0.5662 \\
		VDSH & 0.7165 & 0.7753 & 0.7456 & 0.7318 & 0.6853 & 0.7108 & 0.4410 & 0.5847 & 0.3904 & 0.4327 & 0.1731 & 0.0522 & 0.5366 \\
		NbrReg & n.a. & n.a. & n.a. & n.a. & n.a. & n.a. & n.a. & n.a. & 0.4120 & 0.4644 & 0.4768 & 0.4893 & 0.4249\\
		NASH & 0.7624 & 0.7993 & 0.7812 & 0.7559 & 0.6573 & 0.6921 & 0.6548 & 0.5998 & 0.5108 & 0.5671 & 0.5071 & 0.4664 & 0.6462 \\
		GMSH & 0.7672 & 0.8183 & 0.8212 & 0.7846 & 0.6736 & 0.7024 & 0.7086 & 0.7237 & 0.4855 & 0.5381 & 0.5869 & 0.5583 & 0.6807 \\
		AMMI & 0.8173 & 0.8446 & 0.8506 & 0.8602 & 0.7096 & 0.7416 & 0.7522 & 0.7627 & 0.5518 & 0.5956 & 0.6398 & 0.6618 & 0.7323 \\
		CorrSH & 0.8212 & 0.8420 & 0.8465 & 0.8482 & 0.7243 & 0.7534 & 0.7606 & 0.7632 & \textbf{0.5839} & 0.6183 & 0.6279 & 0.6359 & 0.7355\\
		\midrule
		\Ours{} & \textbf{0.8320} & \textbf{0.8466} & \textbf{0.8560} & \textbf{0.8624} & \textbf{0.7251} & \textbf{0.7543} & \textbf{0.7658} & \textbf{0.7726} & 0.5775 & \textbf{0.6387} & \textbf{0.6646} & \textbf{0.6731} & \textbf{0.7474}\\
		\bottomrule
	\end{tabular}}
	\caption{The precision on three datasets with different numbers of bits in unsupervised document hashing.}
	\label{table:results}
	\vspace{-4mm}
\end{table*}
\subsection{Experiment Setup}
\paragraph{Datasets} We verify the proposed methods on three public datasets which published by VDSH\footnote{https://github.com/unsuthee/VariationalDeepSemantic\\Hashing/tree/master/dataset}: \romannum{1}) {Reuters25178}, which contains 10,788 news documents with 90 different categories;  \romannum{2}) TMC, which is a collection of 21,519 air traffic reports with 22 different categories; \romannum{3}) 20Newsgroups (NG20), which consists of 18,828 news posts from 20 different topics. Note that the category labels of each dataset are only used to compute the evaluation metrics, as we focus on unsupervised scenarios.

\paragraph{Baselines} We compare our method with the following models: SpH \cite{weiss2009spectral}, STH \cite{zhang2010self}, VDSH \cite{chaidaroon2017variational}, NASH \cite{shen2018nash}, GMSH\cite{dong2019document}, NbrReg \cite{chaidaroon2018deep}, CorrSH \cite{zheng2020generative} and AMMI \cite{stratos2020learning}. For all baselines, we take the reported performance from their original papers.

\paragraph{Training Details}
For fair comparisons, we follow the same network architecture used in VDSH, GMSH and CorrSH, using a one-layer feed-forward neural network as the variational and the correlated encoder.
The graph ${\mathbb{G}}$ is constructed with the $K$-nearest neighbors (KNN) algorithm based on cosine similarity on the  TFIDF features of documents. In our experiments, the correlation strength coefficient $\lambda$ in \eqref{correlated_prior} is fixed to 0.99. According to the performance observed on the validation set, we choose the learning rate from $\{0.0005, 0.001, 0.003\}$, batch size from $\{32, 64, 128\}$, the temperature $\tau$ in sigmoid function from $\{0.1, 0.2, \cdots, 1\}$, the number of trees $M$ and neighbors $K$ both form \{1,2,\dots,20\}, with the best used for evaluation on the test set. The model is trained using the Adam optimizer \cite{kingma2014adam}. More detailed experimental settings, along with the generating method of spanning trees, are given in the supplementary materials.

\paragraph{Evaluation Metrics} The retrieval precision is used as our evaluation metric. For each query document, we retrieve 100 documents most similar to it based on the Hamming distance of hash codes. Then, the retrieval precision for a single sample is measured as the percentage of the retrieved documents with the same label as the query. Finally, the average precision over the whole test set is calculated as the performance of the evaluated method.

\subsection{Performance and Analysis}
\begin{table}[!t]
\centering
\small
\setlength{\tabcolsep}{1.5mm}{
    \begin{tabular}{cc|ccccc}
    \toprule
    \multicolumn{2}{c|}{\textbf{Ablation Study}} &  16bits  &  32bits  &  64bits &  128bits \\
    \midrule
    \multirow{2}*{Reuters}
    &  $\text{\Ours}_{\text{ind}}$  &  0.7823   &  0.8094   &  0.8180 &  0.8385 \\
    &  $\text{\Ours}_{\text{prior}}$  &  0.8043   &  0.8295   &  0.8431 &  0.8460 \\
    &  \Ours{}    &  \textbf{0.8320}   &  \textbf{0.8466}   &  \textbf{0.8560} &  \textbf{0.8624} \\
    \midrule
    \multirow{2}*{TMC}
    &  $\text{\Ours}_{\text{ind}}$  &  0.6978   &  0.7307   &  0.7421 &  0.7526 \\
    &  $\text{\Ours}_{\text{prior}}$  &  0.7177  &  0.7408  &  0.7518  &  0.7528  \\
    &  \Ours{}    &  \textbf{0.7251}   &  \textbf{0.7543}   &  \textbf{0.7658} &  \textbf{0.7726} \\
    \midrule
    \multirow{2}*{NG20}
    &  $\text{\Ours}_{\text{ind}}$  &  0.4806   &  0.5503   &  0.6017 &  0.6060 \\
    &  $\text{\Ours}_{\text{prior}}$  &  0.5443   &  0.6071   &  0.6212  & 0.6014 \\
    &  \Ours{}    &  \textbf{0.5775}   &  \textbf{0.6387}   &  \textbf{0.6646} &  \textbf{0.6731} \\
    \bottomrule
    \end{tabular}}
\caption{The performance of variant models. $\text{\Ours}_{\text{ind}}$ and $\text{\Ours}_{\text{prior}}$ indicate the model without considering any document correlations (independent) and only considering correlations in the prior, respectively.
}
\label{table:ablation_study}
\vspace{-4mm}
\end{table}

\paragraph{Overall Performance} The performances of all the models on the three public datasets are shown in Table \ref{table:results}. We see that our model performs favorably to the current state-of-the-art method, yielding best average performance across different datasets and settings.
Compared with VDSH and NASH, which simply employ isotropic Gaussian and Bernoulli prior, respectively, we can observe that our model, which leverages correlated prior and posterior distributions, achieves better results on all the three datasets. Although GMSH improves performance by exploiting a more expressive Gaussian mixture prior, our model still outperforms it by a substantial margin, indicating the superiority of incorporating document correlations.  It is worth noting that, by unifying semantics and neighborhood information under the generative models, the two types of information can be preserved more effectively. This can be validated by that our model performs significantly better than NbrReg, which naively incorporates the neighborhood information by using a neighbor-reconstruction regularizer. The superiority of our unified method can be further corroborated in the comparisons with RBSH and PairRec, which are given in the Supplementary since they employed a different preprocessing method as the models reported here. Comparing to the current SOTA methods of AMMI and CorrSh, our method is still able to achieve better results by exploiting the correlation among documents. Moreover, thanks to the benefit of correlation regularization, remarkable gratuity can be acquired profitably in 64 and 128 bits.
%, although the best previous result given by CorrSH is somewhat higher than the precision of our model in a short bit situation.
\begin{figure}[!t]
	\centering
	\vspace{-2mm}
	\subfigure{
		\begin{minipage}[t]{0.45\linewidth}
			\centering
			\includegraphics[width=1.36in]{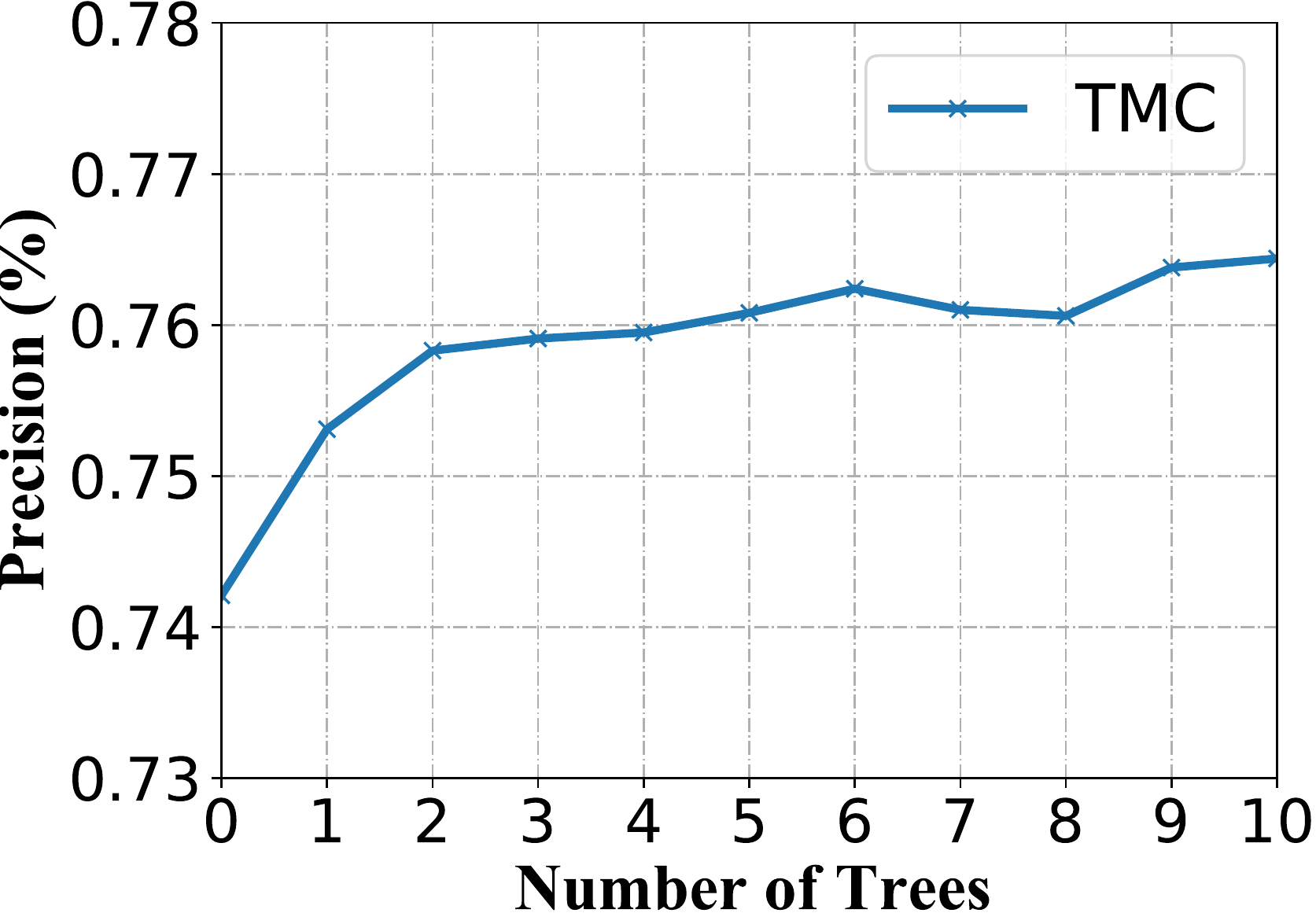}
		\end{minipage}
	}
	\vspace{-2.mm}
	\subfigure{
		\begin{minipage}[t]{0.45\linewidth}
			\centering
			\includegraphics[width=1.36in]{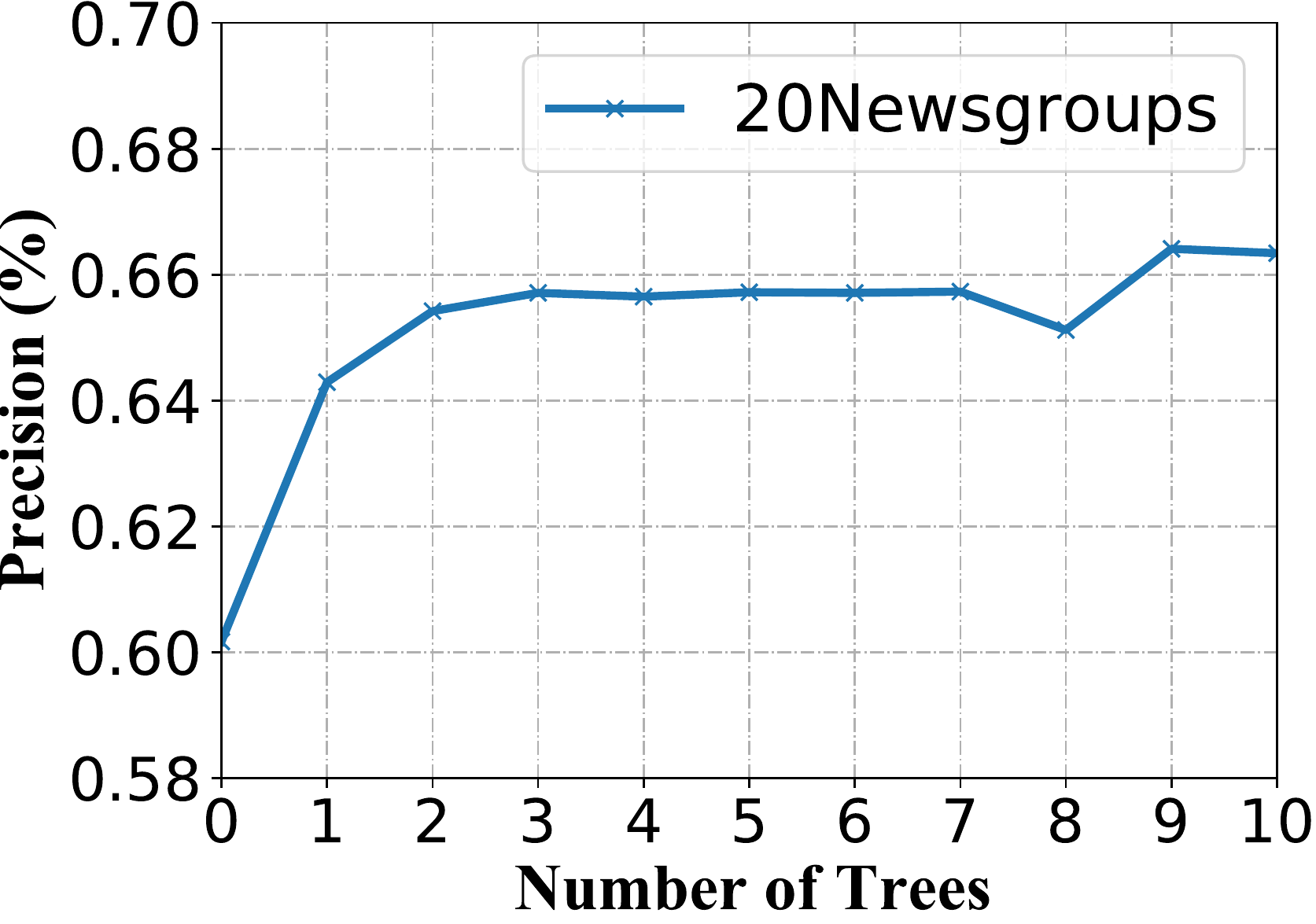}
		\end{minipage}
	}
	\subfigure{
		\begin{minipage}[t]{0.45\linewidth}
			\centering
			\includegraphics[width=1.36in]{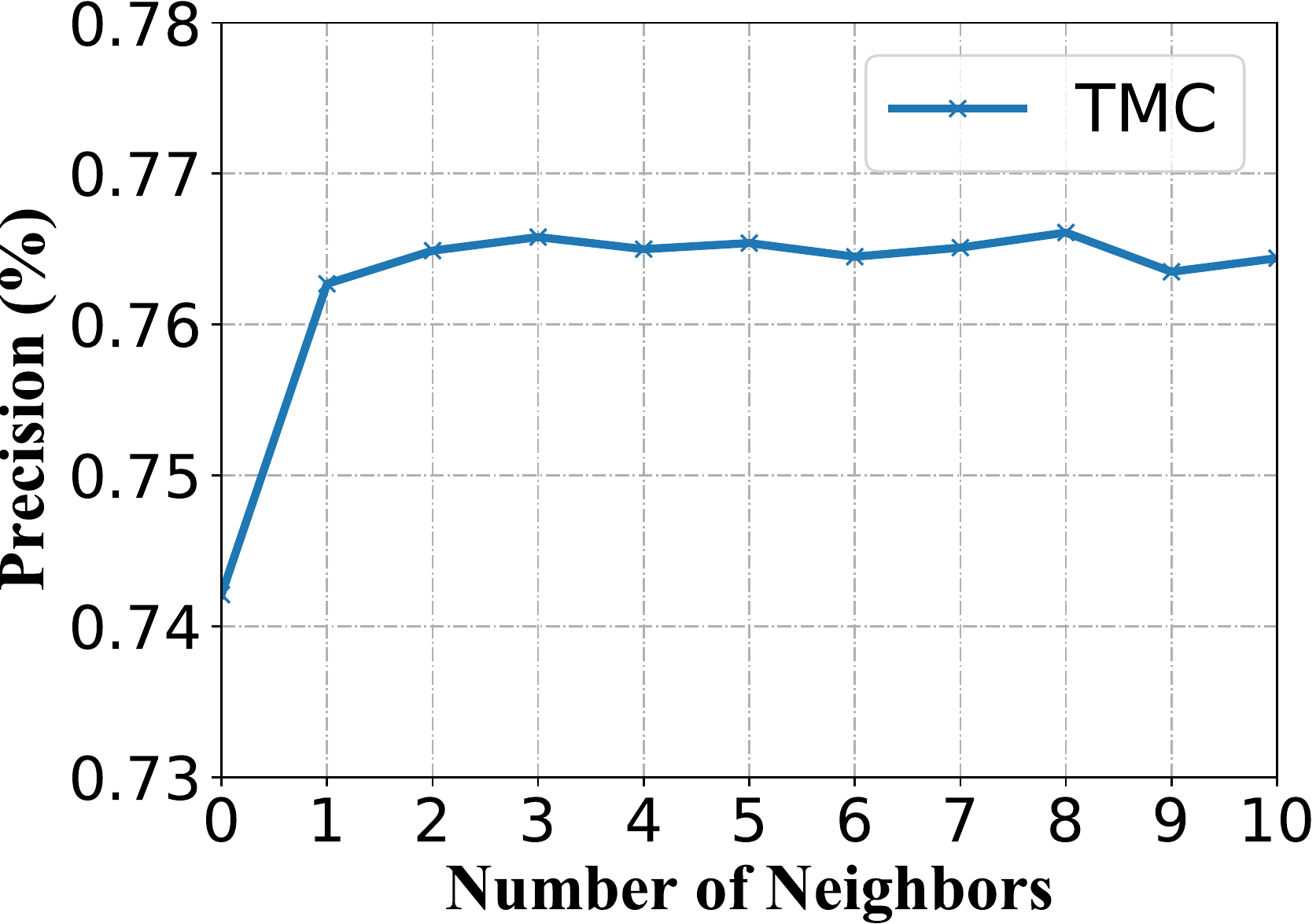}
		\end{minipage}
	}
	\subfigure{
		\begin{minipage}[t]{0.45\linewidth}
			\centering
			\includegraphics[width=1.36in]{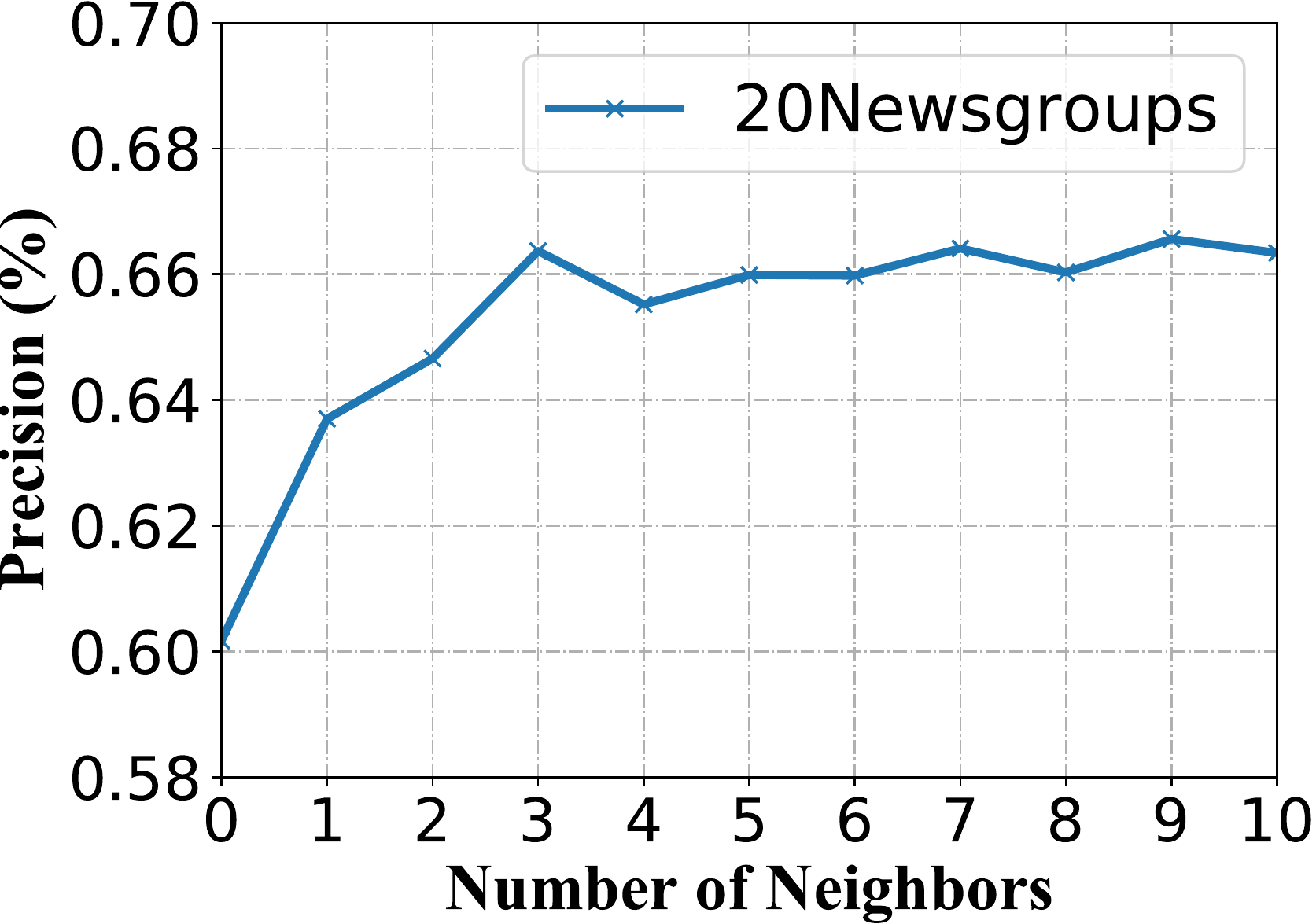}
		\end{minipage}
	}
	\vspace{-2.mm}
	\caption{The precision of 64-bit hash codes with varying number of trees $M$ and neighbors $K$.}
	\label{fig:effect_of_trees}
	\vspace{-4mm}
\end{figure}

\begin{table*}[!t]
\centering
\small
\setlength{\tabcolsep}{1.2mm}{
    \begin{tabular}{c|c|c}
    \toprule
    \textbf{Distance} & \textbf{Category} & \textbf{Title/Subject} \\
    \midrule
    \textbf{query} & \textbf{hockey} & \textbf{NHL PLAYOFF RESULTS FOR GAMES PLAYED 4-21-93} \\
    1 & hockey & NHL PLAYOFF RESULTS FOR GAMES PLAYED 4-19-93 \\
    10 & hockey & NHL Summary parse results for games played Thur, April 15, 1993 \\
    20 & hockey & AHL playoff results (4/15) \\
    50 & forsale & RE:  == MOVING SALE === \\
    70 & hardware & Re: Quadra SCSI Problems? \\
    90 & politics.misc & Re: Employment (was Re: Why not concentrate on child molesters? \\
    \bottomrule
    \end{tabular}}
    \vspace{-1mm}
\caption{Qualitative analysis of the learned 128-bit hash codes on the 20Newsgroups dataset. We present the documents with Hamming distance of 1, 10, 20, 50, 70 and 90 to the query.}
\label{table:case_study}
\end{table*}

\paragraph{Impact of Introducing Correlations in Prior and Posterior} To understand the influences of the proposed document-correlated prior and posterior, we further experiment with two variants of our model: \romannum{1})  $\text{\Ours}_{\text{ind}}$: which does not consider document correlations in neither the prior nor the posterior distribution; \romannum{2}) {$\text{\Ours}_{\text{prior}}$}: which only considers the correlations in the prior, but not in the posterior. Obviously, the proposed $\text{\Ours}$ represents the method that leverage the correlations in both of the prior and posterior. As seen from Table \ref{table:ablation_study}, $\text{\Ours}_{\text{prior}}$ achieves better performance than $\text{\Ours}_{\text{ind}}$, demonstrating the benefit of considering the correlation information of documents only in the prior. By further taking the correlations into account in the posterior, improvements of \Ours{} can be further observed, which fully corroborates the superiority of considering document correlations in the prior and posterior. Another interesting observation is that the performance gap between  $\text{\Ours}_{\text{ind}}$ and {$\text{\Ours}_{\text{prior}}$} becomes small as the length of bits increases. This may be attributed to the fact that the increased generalization ability of models brought by large bits is inclined to alleviate the impact of priori knowledge. However, by additionally incorporating correlation constraints on posterior, significant performance gains would be obtained, especially in large bits scenarios.

\paragraph{Effect of Spanning Trees} For more efficient training, spanning trees are utilized to approximate the whole graph by dropping out some edges. To understand its effects, we first investigate the {\emph{impact of the number of trees}}. The first row of Figure \ref{fig:effect_of_trees}  shows the performance of our method as a function of different numbers of spanning trees. We observe that, compared to not using any correlation, one tree alone can bring significant performance gains. As the tree number increases, the performance rises steadily at first and then converges into a certain level, demonstrating that the document correlations can be mostly captured by several spanning trees. Then, we further explore the {\emph{impact of the neighbor number}} when constructing the graphs using the KNN method, as shown in the second row of Figure \ref{fig:effect_of_trees}. It can be seen that more neighbors contributes to better performance. We hypothesize that this is partly due to the more diverse correlation information captured by the increasing number of neighbors. However, incorporating too many neighbors may lead to the problem of introducing noise and incorrect correlation information to the hash codes. That explains why no further improvement is observed after the number reaches a level.

\paragraph{Empirical Study of Computational Efficiency} We also investigate the training complexity by comparing the training duration of our method and VDSH, on Tesla V100-SXM2-32GB.
On the Reuters, TMC, 20Newsgroups datasets with 64-bit hash codes, our method finishes one epoch of training respectively in 3.791s, 5.238s, 1.343s and VDSH in 2.038s, 4.364s, 1.051s. It can be seen that our model, though with much stronger performance, can be trained almost as efficiently as vanilla VDSH due to the tree approximations.

\paragraph{Case Study} In Table \ref{table:case_study}, we present a retrieval case of the given query document.
It can be observed that as the Hamming distance increases, the semantic (topic) of the retrieved document gradually becomes more irrelevant, illustrating that the Hamming distance can effectively measure the document relevance.

\begin{figure}[!t]
	\centering
	\vspace{-2.5mm}
	\subfigure[\Ours{}]{
		\begin{minipage}[t]{0.4\linewidth}
			\centering
			\includegraphics[width=1.2in]{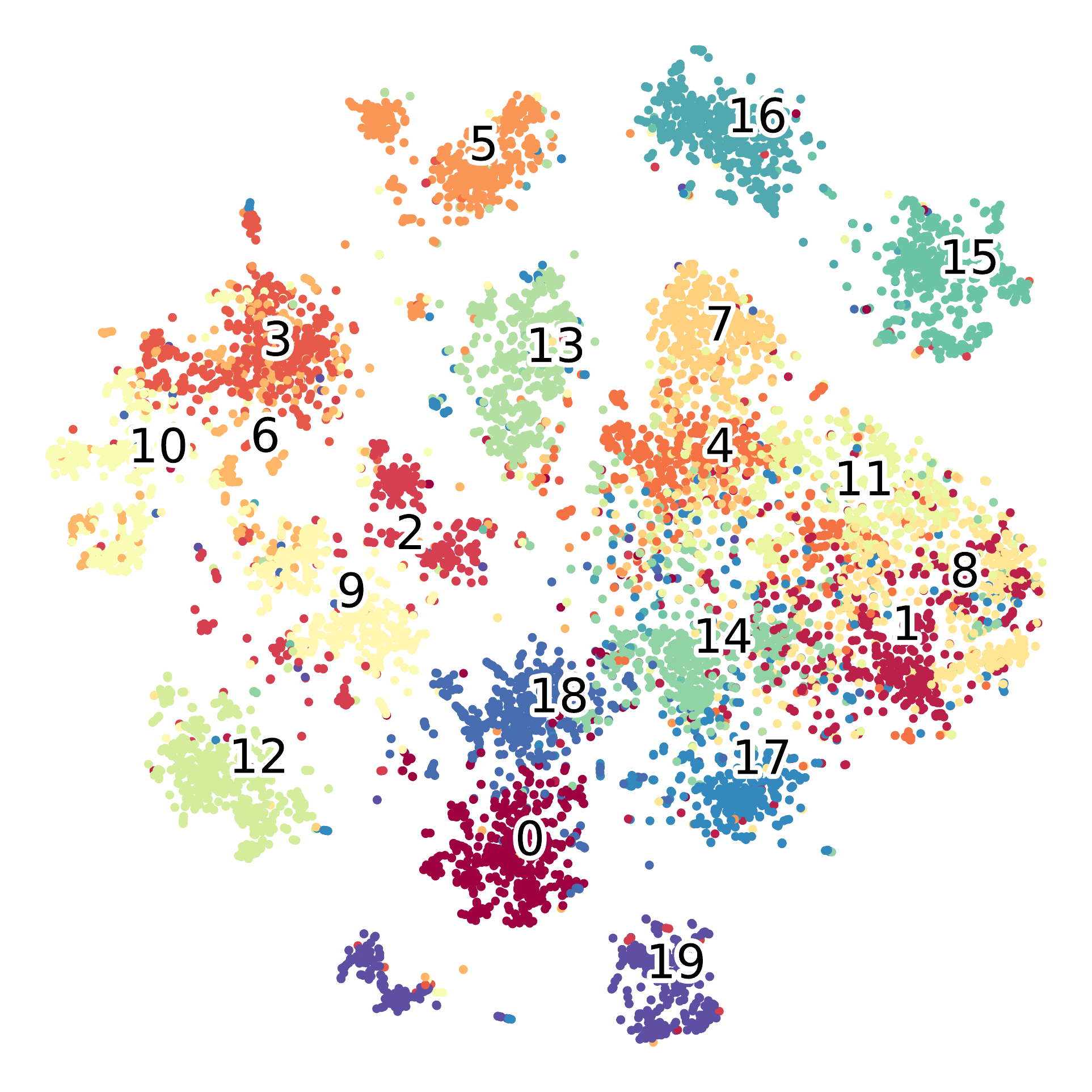}
			%\caption{fig1}
		\end{minipage}
	}
	\vspace{-1.5mm}
	\subfigure[AMMI]{
		\begin{minipage}[t]{0.5\linewidth}
			\centering
			\includegraphics[width=1.5in]{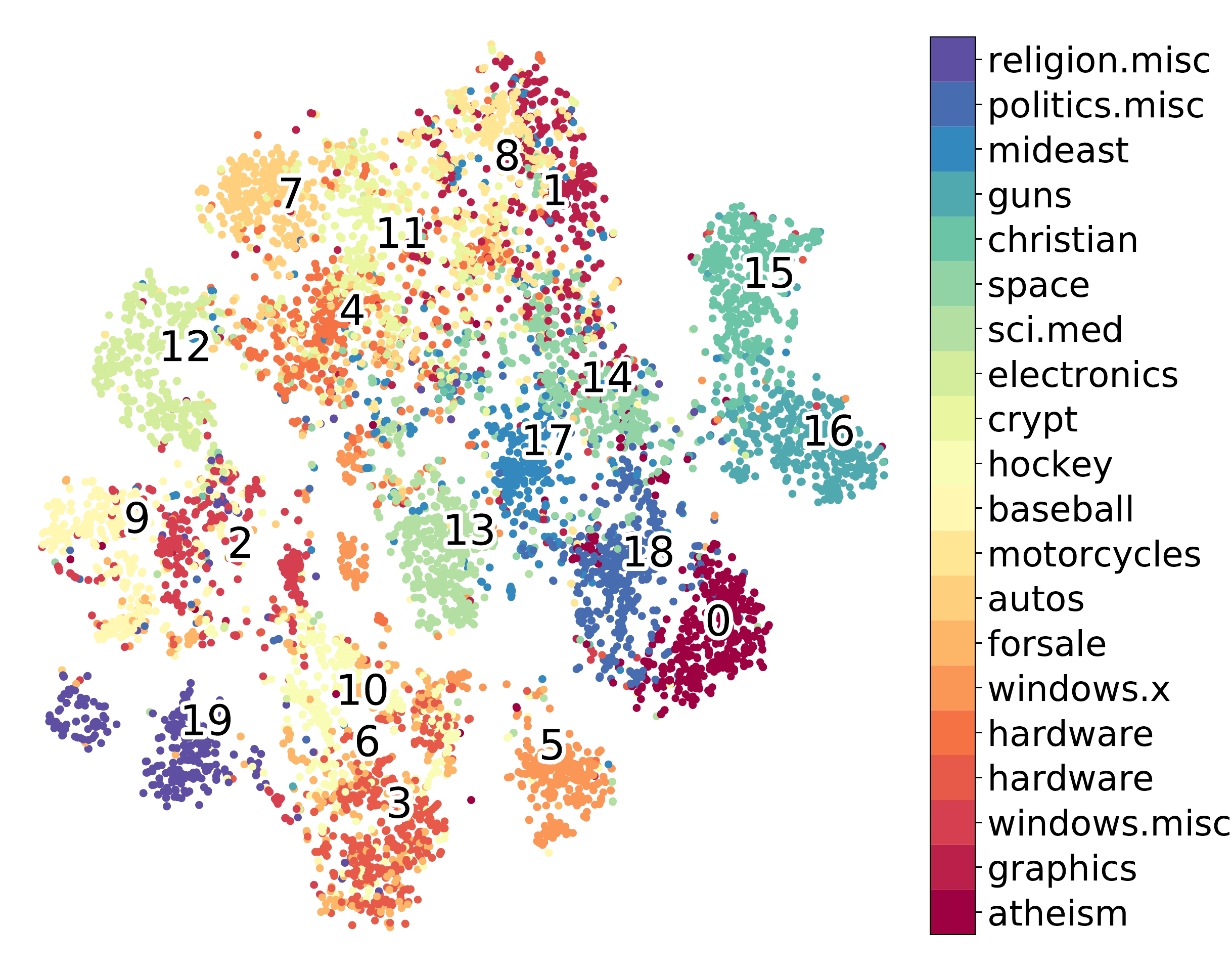}
			%\caption{fig1}
		\end{minipage}
	}
	\vspace{-1.5mm}
	\caption{Visualization of the 64-dimensional latent semantic embeddings learned by the proposed models for the 20Newsgroups dataset.}
	\label{fig:hash_codes_visulization}
	\vspace{-4mm}
\end{figure}

\paragraph{Visualization of Hash Codes} To evaluate the quality of generated hash code more intuitively, we project the latent representations into a 2-dimensional plane with the t-SNE \cite{vanDerMaaten2008} technique. As shown in Figure \ref{fig:hash_codes_visulization}, the representations generated by our method are more separable than those of $\text{AMMI}$, demonstrating the superiority of our method.

\section{Conclusion}
We have proposed an effective and efficient semantic hashing method to preserve both the semantics and neighborhood information of documents. Specifically, we applied a graph-induced Gaussian prior to model the two types of information in a unified framework. To facilitate training, a tree-structure approximation was further developed to decompose the ELBO into terms involving only singleton or pairwise variables. Extensive evaluations demonstrated that our model significantly outperforms baseline methods by incorporating both the semantics and neighborhood information.

\section*{Acknowledgements}
This work is supported by the National Natural Science Foundation of China (No. 61806223, 61906217, U1811264), Key R\&D Program of Guangdong Province (No. 2018B010107005), National Natural Science Foundation of Guangdong Province (No. 2021A1515012299). This work is also supported by MindSpore.

% \section*{Broader Impact}
% In this work, we proposed a novel framework for introducing the correlation of data into binary representations learning. In many real world applications, such as social network analysis, recommendation systems and so on, how to integrate the complex correlation of data into representations is challenging and essential. The ability of the proposed SNUH to learn semantics-neighborhood-unified latent variable models can potentially benefit such applications and therefore benefit the society.

\bibliographystyle{acl_natbib}
\bibliography{acl2021}

\newpage
\appendix
\section*{Appendices}
\section{Derivation of Formulas}

\paragraph{Derivation of $KL\left( q_{\boldsymbol{\phi}}(\boldsymbol{Z} | \boldsymbol{X}) || p_{\mathcal{T}}(\boldsymbol{Z}) \right)$} In the main paper, we propose a tree-type distribution to introduce partial neighborhood information so that the $\mathcal{L}_2$ term can be expressed as the summation over terms involving only one or two variables. Here, we provide the detail derivation.
\begin{equation}
\resizebox{1.01\hsize}{!}{$
\begin{aligned}
    &KL\left( q_{\boldsymbol{\phi}}(\boldsymbol{Z} | \boldsymbol{X}) || p_{\mathcal{T}}(\boldsymbol{Z}) \right) \nonumber \\
    &= \!\! \int \!\!  q_{\boldsymbol{\phi}}(\boldsymbol{Z} | \boldsymbol{X}) \! \log \! \frac{\prod\limits_{i\in {\mathcal{V}}} q_{\boldsymbol{\phi}}(\boldsymbol{z}_i | \boldsymbol{x}_i)}{\prod\limits_{i\in {\mathcal{V}}} \!  p_{\mathcal{G}}(\boldsymbol{z}_i) \!\!\! \prod\limits_{(i, j)\in {\mathcal{E}}_T} \! \frac{p_{\mathcal{G}}(\boldsymbol{z}_i , \boldsymbol{z}_j)}{p_{\mathcal{G}}(\boldsymbol{z}_i) p_{\mathcal{G}}(\boldsymbol{z}_j)}} d\boldsymbol{Z} \nonumber \\
    &= \sum\limits_{i \in \mathcal{V}} KL \left( q_{\boldsymbol{\phi}}(\boldsymbol{z}_i | \boldsymbol{x}_i) || p_{\boldsymbol{\theta}}(\boldsymbol{z}_i) \right) \nonumber \\
    &\!\!\!\quad - \sum\limits_{(i,j) \in \mathcal{E}_T} \!\!\! \mathbb{E}_{q_{\boldsymbol{\phi}}(\boldsymbol{z}_i, \boldsymbol{z}_j | \boldsymbol{x}_i, \boldsymbol{x}_j)} \left[ \log \frac{p_{\mathcal{G}}(\boldsymbol{z}_i) p_{\mathcal{G}}(\boldsymbol{z}_j)}{p_{\mathcal{G}}(\boldsymbol{z}_i, \boldsymbol{z}_j)} \right]. \nonumber
\end{aligned}$}
\end{equation}
Obviously, the KL divergence is decomposed into the terms involving singleton and pairwise variables, which can be calculated efficiently.

\paragraph{Expressing $\mathcal{L}_{\mathcal{T}}$ in Analytical  Form}
 For simplification, in the following, we use $\boldsymbol{\mu}_1, \boldsymbol{\Sigma}_1$ to represent the mean and variance matrix of $q_{\mathcal{T}}(\boldsymbol{z}_i, \boldsymbol{z}_j | \boldsymbol{x}_i, \boldsymbol{x}_j)$, respectively, and represent those of $p_{\mathcal{G}}(\boldsymbol{z}_i, \boldsymbol{z}_j)$ as $\boldsymbol{\mu}_2, \boldsymbol{\Sigma}_2$, respectively. Besides we denote $\lambda a_{ij}$ as $\tau_{ij}$ so we have $\tau_{ij} = \lambda a_{ij} = \lambda a_{ji}$.
 By applying the Cholesky decomposition on the covariance matrix of $\boldsymbol{\Sigma}_1$ and $\boldsymbol{\Sigma}_2$
 \begin{align}
 \boldsymbol{\Sigma}_1 \!\! &= \!\! \left[ \!\! \begin{array}{cc}
    \boldsymbol{\sigma}_{i} \!&\! \boldsymbol{0}_{d} \\
    \boldsymbol{\gamma}_{ij} \boldsymbol{\sigma}_{j} \!&\! \sqrt{1 - \boldsymbol{\gamma}_{ij}^2 } \boldsymbol{\sigma}_{j}
    \end{array} \!\!\! \right] \!\!\! \left[ \!\! \begin{array}{cc}
    \boldsymbol{\sigma}_{i} \!&\! \boldsymbol{\gamma}_{ij} \boldsymbol{\sigma}_{j} \\
     \boldsymbol{0}_{d} \!&\! \sqrt{1 - \boldsymbol{\gamma}_{ij}^2 } \boldsymbol{\sigma}_{j}
    \end{array} \!\! \! \right], \nonumber \\
\boldsymbol{\Sigma}_{2} \! &= \!\!\left[\begin{array}{cc}
    \boldsymbol{I}_{d} \!\!&\!\! 0 \\
    \tau_{ij} \boldsymbol{I}_{d} \!\!&\!\! \sqrt{1 - \tau_{ij}^{2}} \boldsymbol{I}_{d}
    \end{array}\right] \!\!\! \left[\begin{array}{cc}
    \boldsymbol{I}_{d} \!\!&\!\! \tau_{ij} \boldsymbol{I}_{d} \\
    0 \!\!&\!\! \sqrt{1 - \tau_{ij}^{2}} \boldsymbol{I}_{d}
    \end{array}\right], \nonumber 
\end{align}
where we omit $diag(\cdot)$ for simplifying, we have 
\begin{equation}
\resizebox{1.0\hsize}{!}{$
\begin{aligned}
&{KL}\left(q_{\boldsymbol{\phi}}(\boldsymbol{z}_i, \boldsymbol{z}_j | \boldsymbol{x}_i, \boldsymbol{x}_j) || p_{\mathcal{G}}(\boldsymbol{z}_i, \boldsymbol{z}_j)\right) \\
&\!\!= \frac{1}{2} \sum\limits_{n=1}^d \Big\{ \log (1-\tau_{ij}^2) \\
&\!\!- \big( \log \boldsymbol{\sigma}_{in}^2 + \log \boldsymbol{\sigma}_{jn}^2 + \log (1 - \boldsymbol{\gamma}_{ijn}^2)\big) - 2 \\
&\!\! + \! \frac{\boldsymbol{\sigma}_{in}^2 \!\! + \! \boldsymbol{\sigma}_{jn}^2 \!\! - \! 2\tau_{ij} \boldsymbol{\gamma}_{ijn} \boldsymbol{\sigma}_{in} \boldsymbol{\sigma}_{jn} \!\! + \! \boldsymbol{\mu}_{in}^2 \!\! + \! \boldsymbol{\mu}_{jn} \!\! - \! 2\tau_{ij} \boldsymbol{\mu}_{in} \boldsymbol{\mu}_{jn}}{1-\tau_{ij}^2} \!\!  \Big\}. \nonumber
\end{aligned}$}
\end{equation}
Then, we can express $\mathcal{L}_{\mathcal{T}}$ in an analytical form
\begin{equation}
\resizebox{1.0\hsize}{!}{$
\begin{aligned}
&\mathcal{L}_{\mathcal{T}} = \sum\limits_{i \in \mathcal{V}} \Big( \mathbb{E}_{q_{\boldsymbol{\phi}}(\boldsymbol{z}_i | \boldsymbol{x}_i)} [\log p_{\boldsymbol{\theta}}(\boldsymbol{x}_i | \boldsymbol{z}_i)] - \frac{1}{2} \sum\limits_{n=1}^d (\boldsymbol{\mu}_{in}^2 \\
&\!\! + \boldsymbol{\sigma}_{in}^2 \! - \! 1 \! - \! 2\log \boldsymbol{\sigma}_{in}) \Big) \! - \!\!\!\!\! \sum\limits_{(i,j) \in \mathcal{E}_{T}} \!\!\! \bigg( \frac{1}{2} \! \sum\limits_{n=1}^d \! \Big\{ \! \log (1-\tau_{ij}^2) \\
&\!\! - \big( \boldsymbol{\mu}_{in}^2 + \boldsymbol{\mu}_{jn}^2 + \boldsymbol{\sigma}_{in}^2 + \boldsymbol{\sigma}_{jn}^2 + \log (1 - \boldsymbol{\gamma}_{ijn}^2)\big) \\
&\!\! + \! \frac{\boldsymbol{\sigma}_{in}^2 \!\! + \! \boldsymbol{\sigma}_{jn}^2 \!\!\! - \! 2\tau_{ij} \boldsymbol{\gamma}_{ijn} \boldsymbol{\sigma}_{in} \boldsymbol{\sigma}_{jn} \!\! + \! \boldsymbol{\mu}_{in}^2 \!\! + \! \boldsymbol{\mu}_{jn} \!\!\! - \! 2\tau_{ij} \boldsymbol{\mu}_{in} \boldsymbol{\mu}_{jn}}{1-\tau_{ij}^2} \!\! \Big\} \!\! \bigg) \nonumber
\end{aligned}$}
\end{equation}

\begin{algorithm}[!t]
\caption{Model Training Algorithm}
\label{alg:training_alg}
\small
\textbf{Input}: Document representations $\boldsymbol{X}$; edges list of spanning trees $\boldsymbol{E}$; batch size $b$.\\
\textbf{Output}: Optimal parameters ($\boldsymbol{\theta}, \boldsymbol{\phi}$).
\begin{algorithmic}[1] %[1] enables line numbers
    \State $\boldsymbol{\theta}, \boldsymbol{\phi} \leftarrow$ Initialize parameters
    \Repeat
        \State $\mathcal{V}^{M} \!\leftarrow\! \{\boldsymbol{x}_1, \cdots, \boldsymbol{x}_b \} \!\sim\! \boldsymbol{X}$ \Comment{Sample nodes}
        \State $\mathcal{E}_{T}^{M} \!\leftarrow\! \{e_1, \cdots, e_b \} \!\sim\! \boldsymbol{E}$ \Comment{Sample endges}
        \State $\boldsymbol{g} \leftarrow \nabla_{\boldsymbol{\phi}, \boldsymbol{\theta}} \mathcal{\widetilde L }_{\mathcal{MT}}^{M}(\boldsymbol{\theta}, \boldsymbol{\phi}; \mathcal{V}^{M}, \mathcal{E}_{T}^{M})$
        \State $\boldsymbol{\theta}, \boldsymbol{\phi} \leftarrow$ Update parameters using gradients $\boldsymbol{g}$ ({\it e.g.}, Adam optimizer)
    \Until convergence of parameters ($\boldsymbol{\theta}, \boldsymbol{\phi}$)
\end{algorithmic}
\end{algorithm}

\begin{table}[!t]
\vspace{-4mm}
\centering
\small
\setlength{\tabcolsep}{.8mm}{
    \begin{tabular}{c|cccc}
    \toprule
    Input & \multicolumn{4}{c}{document pair ($\boldsymbol{x}_i$; $\boldsymbol{x}_j$)} \\
    \midrule
    & \multicolumn{2}{c|}{\textbf{Variational Enc}} & \multicolumn{2}{c}{\textbf{Correlated Enc}} \\
    \midrule
    \multirow{2}*{Encoder}
    & Linear($|V|, d$) & Linear($|V|, d$) & \multicolumn{2}{|c}{Linear($2 |V|, d$)}  \\
    & $\boldsymbol{\mu} = f(\cdot / \tau)$ & $\boldsymbol{\sigma} = g(\cdot)$ & \multicolumn{2}{c}{$\boldsymbol{\gamma} = 2 * f(\cdot) - 1$} \\
    \midrule
    Generator & \multicolumn{4}{c}{Linear($d, |V|$)} \\
    \bottomrule
    \end{tabular}}
\caption{The neural network architecture of the proposed model, in which $f(\cdot)$ and $g(\cdot)$ represent the sigmoid and softplus function, respectively.}
\label{table:network_architecture}
\vspace{-4mm}
\end{table}

\paragraph{Derivation of $\mathcal{\widetilde L }_{\mathcal{MT}}$} With $\mathcal{L}_{\mathcal{MT}}$, we extend the single-tree approximation to multi-tree approximation. Although the KL divergence between the mixture distributions does not have a closed-form solution, we can obtain its explicit upper bound by using the log-sum inequality as
\begin{align}
    \mathcal{L}_{\mathcal{MT}} 
    &\geq \frac{1}{M} \sum\limits_{\mathcal{T} \in \mathcal{T}_{\mathbb{G}}} \mathbb{E}_{q_{\mathcal{T}}(\boldsymbol{Z} | \boldsymbol{X})}[\log p_{\boldsymbol{\theta}}(\boldsymbol{X} | \boldsymbol{Z})] \nonumber \\
    &\qquad\quad \!\!\!\!\!\! - \frac{1}{M} \sum\limits_{\mathcal{T} \in \mathcal{T}_{\mathbb{G}}}  KL\left( q_{\mathcal{T}}(\boldsymbol{Z} | \boldsymbol{X}) || p_{\mathcal{T}}(\boldsymbol{X}) \right) \nonumber \\
    &\triangleq \mathcal{\widetilde L }_{\mathcal{MT}}. \nonumber
\end{align}
We can further express $\mathcal{\widetilde L }_{\mathcal{MT}}$ in a more intuitive form as
\begin{equation}
\resizebox{1.02\hsize}{!}{$
\begin{aligned}
    &\sum\limits_{i \in \mathcal{V}} \!\! \Big( \mathbb{E}_{q_{\boldsymbol{\phi}}(\boldsymbol{z}_i | \boldsymbol{x}_i)}\![\log p_{\boldsymbol{\theta}}(\boldsymbol{x}_i | \boldsymbol{z}_i)] \!\! - \!\! KL \!\left( q_{\boldsymbol{\phi}}(  \boldsymbol{z}_i | \boldsymbol{x}_i) || p_{\mathcal{G}}(\boldsymbol{z}_i) \right) \!\! \Big) \nonumber \\
    &\!\! - \sum\limits_{(i,j) \in \mathcal{E}_{T}} \!\!\! w_{ij} \Big( KL \! \left( q_{\boldsymbol{\phi}}(\boldsymbol{z}_i, \boldsymbol{z}_j | \boldsymbol{x}_i, \boldsymbol{x}_j) || p_{\mathcal{G}}(\boldsymbol{x}_i, \boldsymbol{x}_j) \right) \nonumber \\
    &\!\! - \! KL \! \left( q_{\boldsymbol{\phi}}(\boldsymbol{z}_i | \boldsymbol{x}_i) || p_{\mathcal{G}}(\boldsymbol{z}_i)\right) \!\! - \!\! KL \! \left( q_{\boldsymbol{\phi}}(\boldsymbol{z}_j | \boldsymbol{x}_j) || p_{\mathcal{G}}(\boldsymbol{z}_j)\right) \!\! \Big), \nonumber
\end{aligned}$}
\end{equation}
where $w_{ij} = \frac{\left|\{{\mathcal{T}} \in \mathcal{T}_{\mathbb{G}} | (i, j) \in \mathcal{E}_T \} \right|}{M}$ denotes the proportion of times that the edge $(i,j)$ appears. To optimize this objective, we can construct an estimator of the ELBO, based on the minibatch
\begin{align}
    \mathcal{\widetilde L }_{\mathcal{MT}}
    &\simeq  \mathcal{\widetilde L }_{\mathcal{MT}}^{M} \nonumber \\
    &= \!\! \sum\limits_{i \in \mathcal{V}^{M}} \mathcal{L}_{\mathcal{V}^{M}}(\boldsymbol{x}_i) - \!\!\!\!\! \sum\limits_{(i,j) \in \mathcal{E}_{T}^{M}} \!\!\!\! w_{ij} \mathcal{L}_{\mathcal{E}_{T}^{M}}(\boldsymbol{x}_i, \boldsymbol{x}_j), \nonumber
\end{align}
where $\mathcal{V}^{M}$ is the subset of documents, $\mathcal{E}_{T}^{M}$ is the subset of edges and
\begin{align}
    \mathcal{L}_{\mathcal{V}^{M}}(\boldsymbol{x}_i) \triangleq \mathbb{E}_{q_{\boldsymbol{\phi}}(\boldsymbol{z}_i | \boldsymbol{x}_i)} &[\log p_{\boldsymbol{\theta}}(\boldsymbol{x}_i | \boldsymbol{z}_i)]  \nonumber \\
    & \!\!\!\!\!\!\!-  KL \left( q_{\boldsymbol{\phi}}(  \boldsymbol{z}_i | \boldsymbol{x}_i) || p_{\mathcal{G}}(\boldsymbol{z}_i) \right); \nonumber
\end{align}
\vspace{-10.mm}
\begin{align}
    &\mathcal{L}_{\mathcal{E}_{T}^{M}}(\boldsymbol{x}_i, \boldsymbol{x}_j) \! \triangleq \! KL \! \left( q_{\boldsymbol{\phi}}(\boldsymbol{z}_i, \boldsymbol{z}_j | \boldsymbol{x}_i, \boldsymbol{x}_j) || p_{\mathcal{G}}(\boldsymbol{x}_i, \boldsymbol{x}_j) \right) \nonumber \\
    &\!\!-\!\! K\!L \! \left( q_{\boldsymbol{\phi}}(\boldsymbol{z}_i | \boldsymbol{x}_i) || p_{\mathcal{G}}(\boldsymbol{z}_i)\right) \! - \!\! K\!L \!\! \left( q_{\boldsymbol{\phi}}(\boldsymbol{z}_j | \boldsymbol{x}_j) || p_{\mathcal{G}}(\boldsymbol{z}_j)\right). \nonumber
\end{align}
Then we can update the parameters by using the gradient $\nabla_{\boldsymbol{\phi}, \boldsymbol{\theta}} \mathcal{\widetilde L }_{\mathcal{MT}}^{M}$. The training procedure is summarized in Algorithm \ref{alg:training_alg}.

\begin{algorithm}[!t]
\caption{Spanning Tree Generation Algorithm}
\label{alg:tree_generation}
\small
\textbf{Input}: Graph $\mathbb{G}$; number of trees $n$.\\
% \textbf{Parameter}: Optional list of parameters\\
\textbf{Output}: Edges list of spanning trees $\boldsymbol{E}$.
\begin{algorithmic}[1] %[1] enables line numbers
\Procedure{TreeGen}{$n$}    \Comment{Input: \#tree $n$}
    \State $\boldsymbol{E} = [\ ]$ \Comment{Initial edges list}
    \For{$k \leftarrow 0, \cdots, n-1$}
        \State $\boldsymbol{V} = [False]^{|\mathcal{V}|}$ \Comment{Visited node list}
        \While{$False$ in $\boldsymbol{V}$}
            \State $i \leftarrow RC_{[V==False]}$ \Comment{Choose node}
            \State $\boldsymbol{Q} = [i]$    \Comment{Initial queue}
            \While{$len(\boldsymbol{Q}) > 0$}
                \State $i \leftarrow \boldsymbol{Q}[0]$
                \State $\boldsymbol{V}[i] \leftarrow True$
                \State $\boldsymbol{N} = ID_{[V[\mathcal{N}(i)]==False]}$
                \If{$len(\boldsymbol{N}) == 0$}
                    \State $POP(\boldsymbol{Q}, -1)$
                    \State break
                \EndIf
                \State $j \! \leftarrow \! RC_{[\boldsymbol{N}]}$  \Comment{Choose neighbor}
                \State $\boldsymbol{V}[j] \leftarrow True$
                \State $APPEND(\boldsymbol{Q}, j)$
                \State $APPEND(\boldsymbol{E}, [i, j])$
            \EndWhile
        \EndWhile
    \EndFor
\EndProcedure
\end{algorithmic}
\end{algorithm}

% \begin{table*}[ht]
% \centering
% \small
% \setlength{\tabcolsep}{2.0mm}{
%     \begin{tabular}{c|cccccc}
%     \toprule
%     Word & \textbf{weapons} &  \textbf{medical}  &  \textbf{companies}  &  \textbf{define} &  \textbf{israel} & \textbf{book} \\
%     \midrule
%     \multirow{5}*{NASH} 
%     & gun & treatment & company & definition & israeli & books \\
%     & guns & disease & market & defined & arabs & english \\
%     & weapon & drugs & afford & explained & arab & references \\
%     & armed & health & products & discussion & jewish & learning \\
%     & assault & medicine & money & knowledge & jews & reference \\
%     \midrule
%     \multirow{5}*{\Ours} 
%     & weapon & medicine & inexpensive & defined & israeli & books \\
%     & armed & disease & expensive & definitions & arab & reference \\
%     & concealed & patients & cost & defines & arabs & chapter \\
%     & guns & physician & manufacturers & definition & palestinian & guide \\
%     & gun & treatment & design & arbitrary & gaza & origin \\
%     \bottomrule
%     \end{tabular}}
% \caption{The five nearest words in the semantic space learned by our model, compared with NASH.}
% \label{table:semantic_information}
% \end{table*}

\begin{table}[ht]
\centering
\small
\setlength{\tabcolsep}{1.5mm}{
    \begin{tabular}{ccccccc}
    \toprule
    Datasets & Methods &  16bits  &  32bits  &  64bits &  128bits \\
    \midrule
    \multirow{3}*{Reuters}    
    &  RBSH  &  0.7740   &  0.8149   &  0.8120 &  0.8088 \\
    &  PairRec  &  0.8028   &  0.8268   &  0.8329 &  0.8468 \\
    &  \Ours{}    &  \textbf{0.8063}   &  \textbf{0.8369}   &  \textbf{0.8483} &  \textbf{0.8567} \\
    \midrule
    \multirow{3}*{TMC}        
    &  RBSH  &  0.7959  &  0.8138  &  0.8224  &  0.8193  \\
    &  PairRec  &  0.7991  &  0.8239  &  0.8280  &  0.8303  \\
    &  \Ours{}    &  {0.7901}   &  {0.8145}   &  \textbf{0.8293} &  \textbf{0.8329} \\
    \midrule
    \multirow{3}*{NG20}
    &  RBSH  &  0.6087   &  0.6385   &  0.6655  & 0.6668 \\
    &  PairRec  &   n.a.  &  n.a.   &  n.a.  & n.a. \\
    &  \Ours{}    &  {0.5679}   &  \textbf{0.6444}   &  \textbf{0.6806} &  \textbf{0.7004} \\
    \bottomrule
    \end{tabular}}
\caption{The precision of variant models on three datasets with different numbers of bits.}
\label{table:compare_RBSH}
\vspace{-4mm}
\end{table}
\section{Tree Generation Algorithm}
Algorithm \ref{alg:tree_generation} shows the spanning tree generation algorithm \text{TreeGen($\cdot$)} used in our graph-induced generative document hashing model. \text{TreeGen($\cdot$)} utilizes a depth-first search (DFS) algorithm to generate meaningful neighborhood information for each node. In this algorithm, $RC_{[\cdot]}$ means randomly choosing one index according to the indicator function; $ID_{[\cdot]}$ represents the set of node indexes satisfying the indicator condition and $\mathcal{N}(i)$ denotes the neighbors of node $i$. Due to the importance of edges precision, when choosing a neighbor (line 16 in Algorithm \ref{alg:tree_generation}), instead of using uniform sampling, we exploit a temperature $\alpha$ to control the trade-off between the precision and diversity of edges. Specifically, the probability of sampling neighbor $j$ of node $i$ is
\begin{align}
    \frac{exp(cos(\boldsymbol{x}_j^T \boldsymbol{x}_i) / \alpha)}{\sum_{n \in \mathcal{N}(i)} exp(cos(\boldsymbol{x}_n^T \boldsymbol{x}_i) / \alpha)}. \nonumber
\end{align}
We find the best configuration of $\alpha$ on the validation set with the values in $\{0.1, 0.2, \cdots, 1\}$ .

\section{Experiment Details} For fair comparisons, we follow the experimental setting of VDSH. Specifically, the vocabulary size $|V|$ is 7164, 20000, and 10000 for Reuters, TMC and 20Newsgroups, respectively. The split of training, validation, and test set is as follows: 7752, 967, 964 for  Reuters; 21286, 3498, 3498 for TMC and 11016, 3667, 3668 for 20Newsgroups, respectively. Moreover, the KL term in Eq. (18) of the main paper is weighted with a coefficient $\beta$ to avoid posterior collapse. We find the best conﬁguration of $\beta$ on the validation set with the values in $\{0.01, 0.02, \cdots, 0.1\}$. To intuitively understand our model, we illustrate the whole architecture in Table \ref{table:network_architecture}.

\begin{figure}[!t]
	\centering
	\vspace{-2mm}
	\subfigure{
		\begin{minipage}[t]{0.45\linewidth}
			\centering
			\includegraphics[width=1.36in]{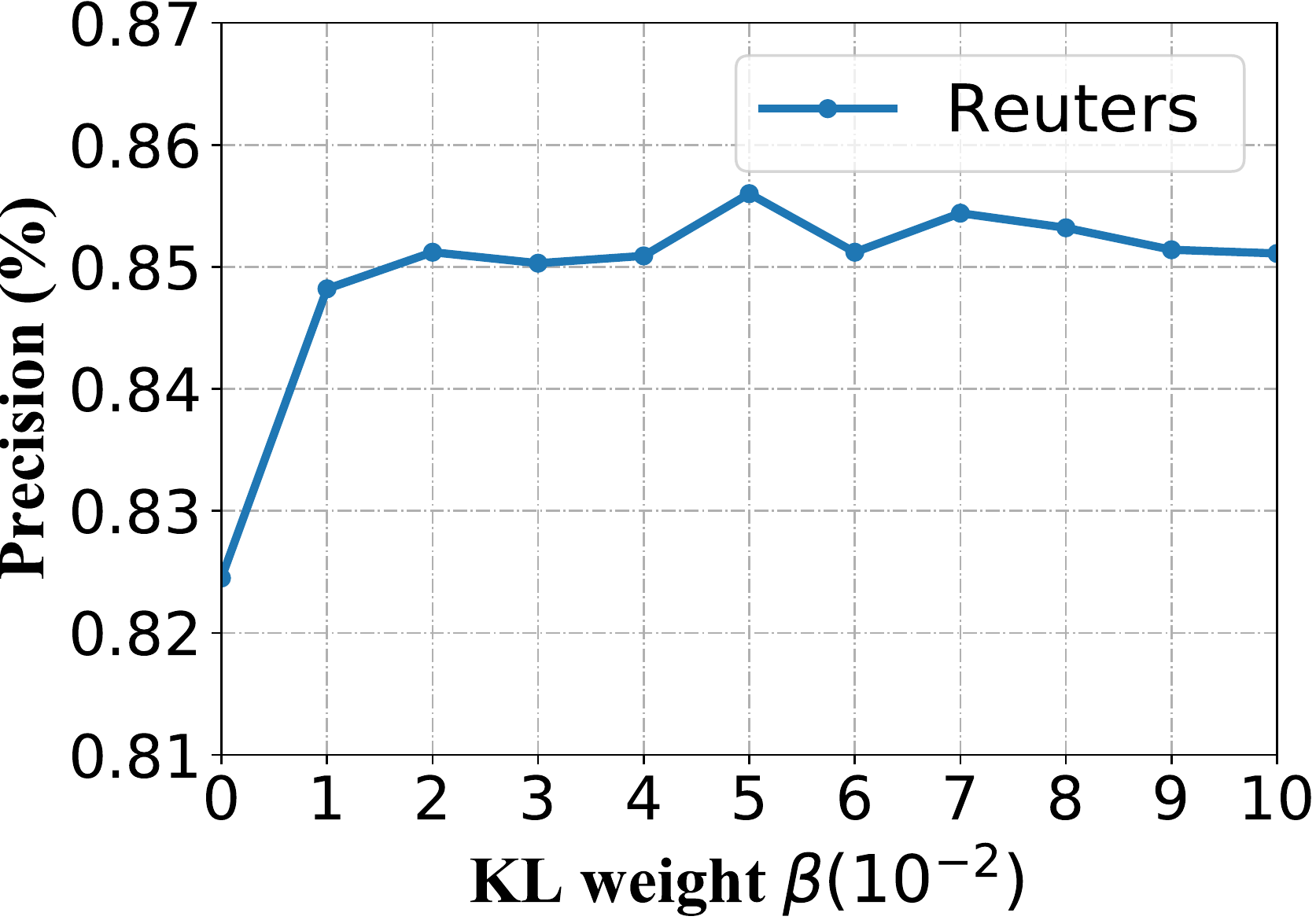}
		\end{minipage}
	}
	\vspace{-2mm}
	\subfigure{
		\begin{minipage}[t]{0.45\linewidth}
			\centering
			\includegraphics[width=1.36in]{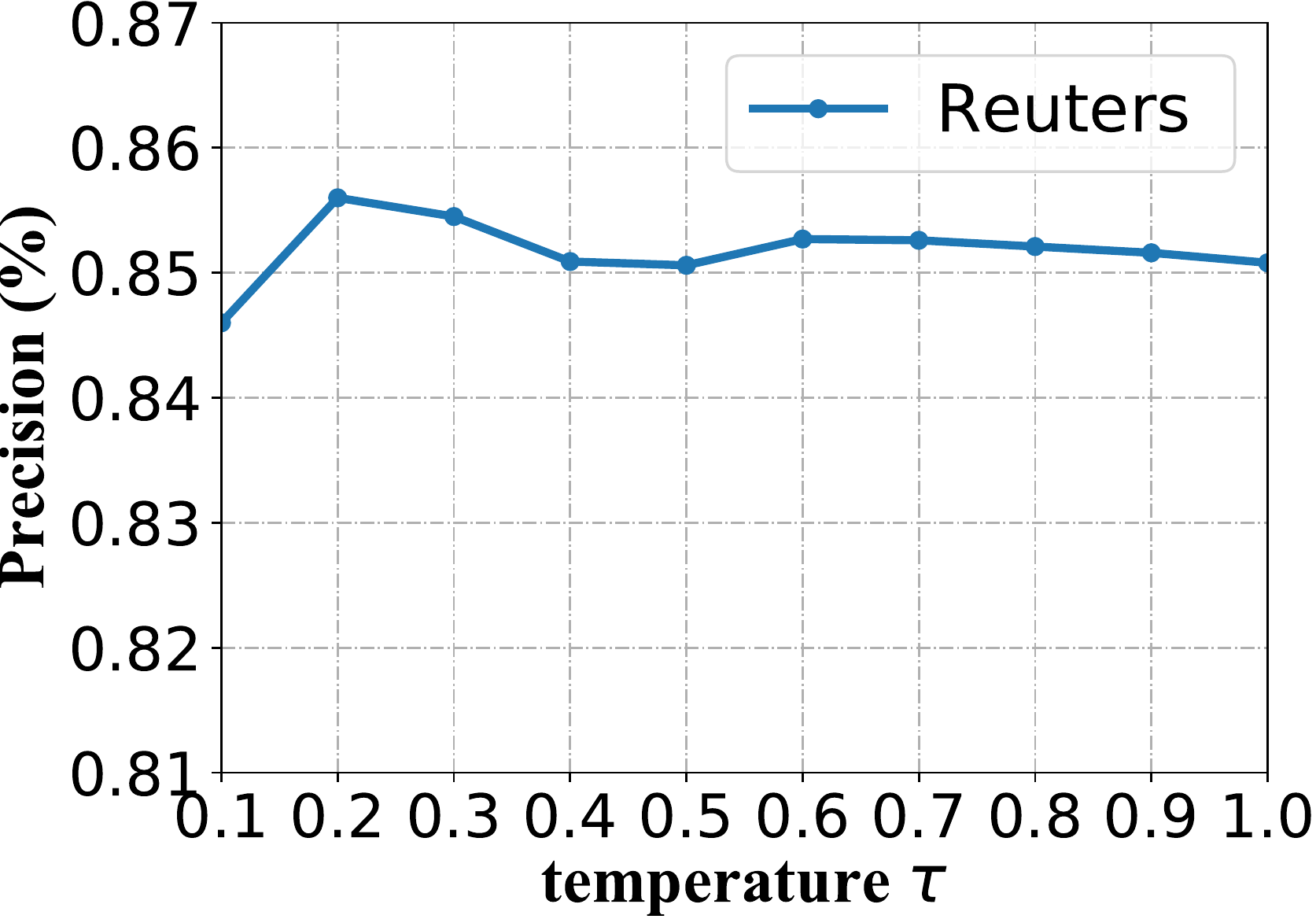}
		\end{minipage}
	}
	\subfigure{
		\begin{minipage}[t]{0.45\linewidth}
			\centering
			\includegraphics[width=1.36in]{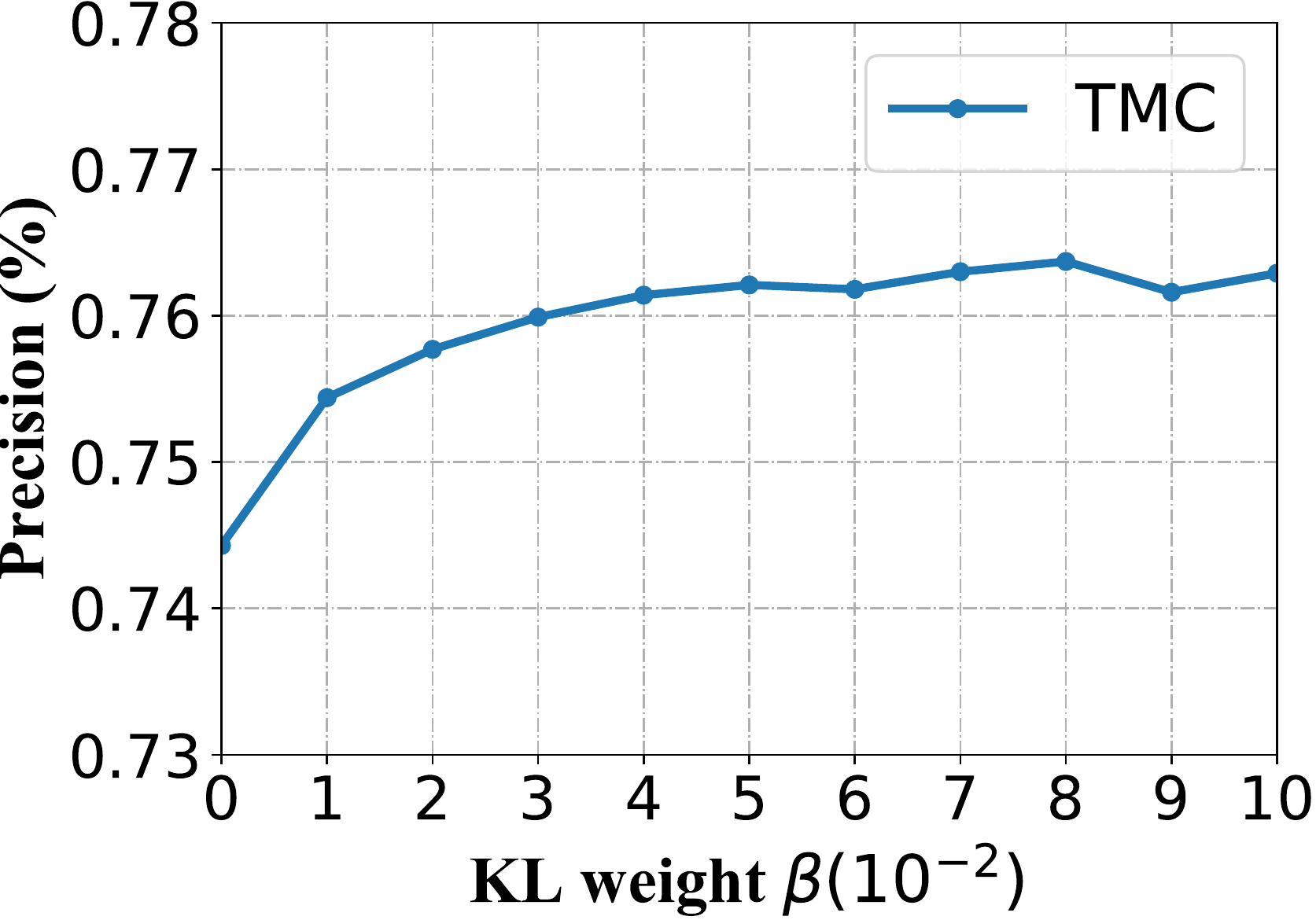}
		\end{minipage}
	}
	\vspace{-2mm}
	\subfigure{
		\begin{minipage}[t]{0.45\linewidth}
			\centering
			\includegraphics[width=1.36in]{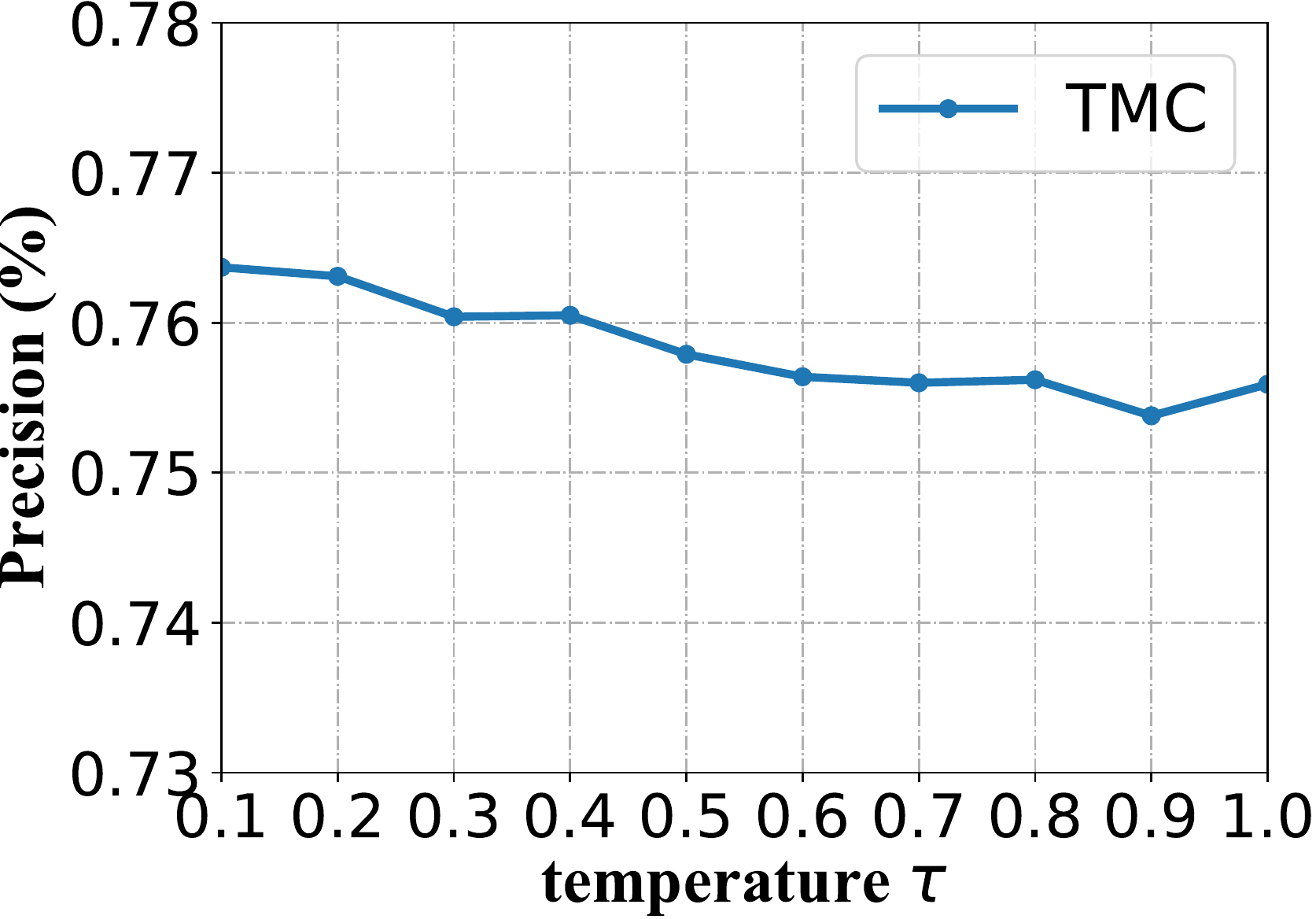}
		\end{minipage}
	}
	\subfigure{
		\begin{minipage}[t]{0.45\linewidth}
			\centering
			\includegraphics[width=1.36in]{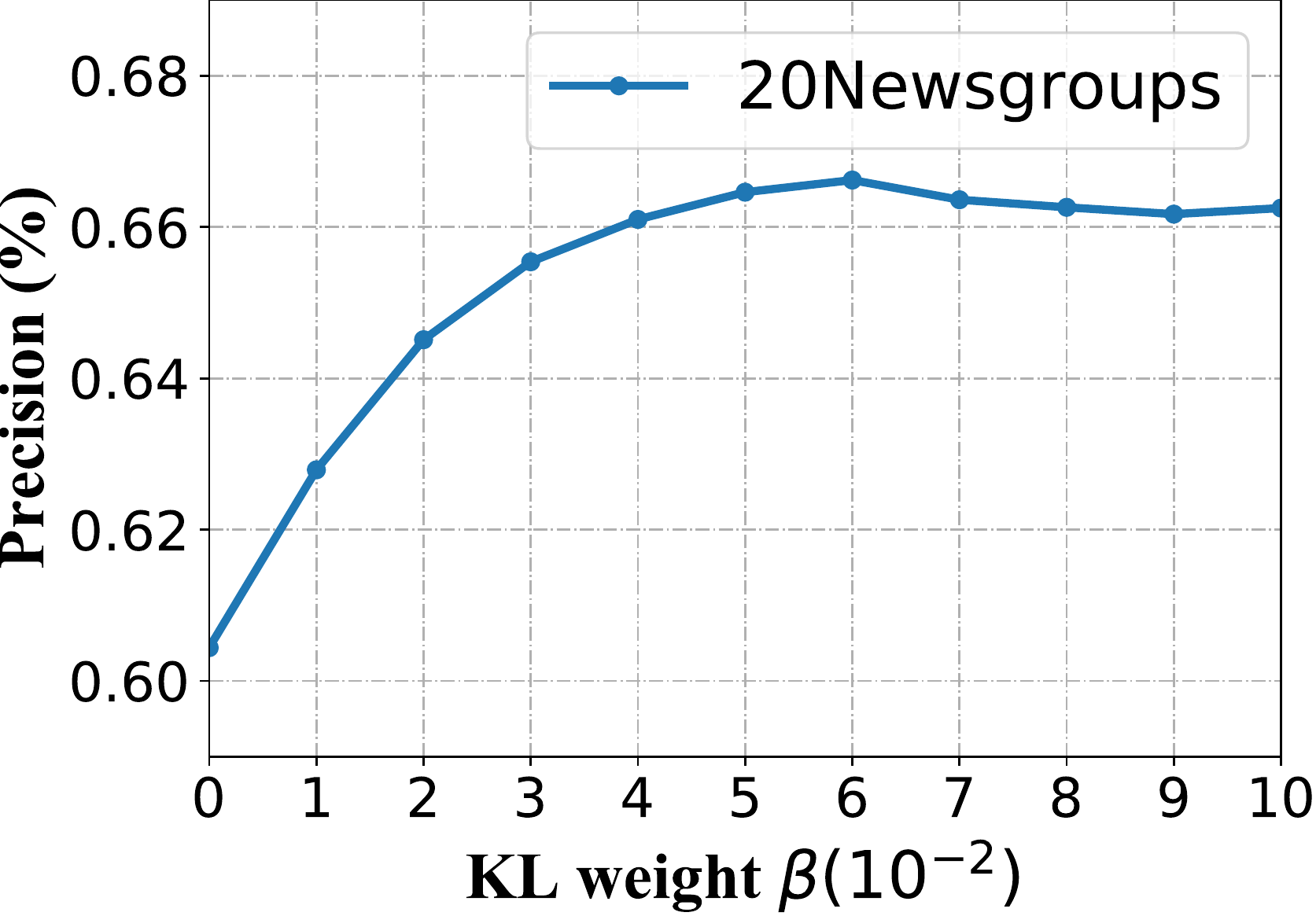}
		\end{minipage}
	}
	\subfigure{
		\begin{minipage}[t]{0.45\linewidth}
			\centering
			\includegraphics[width=1.36in]{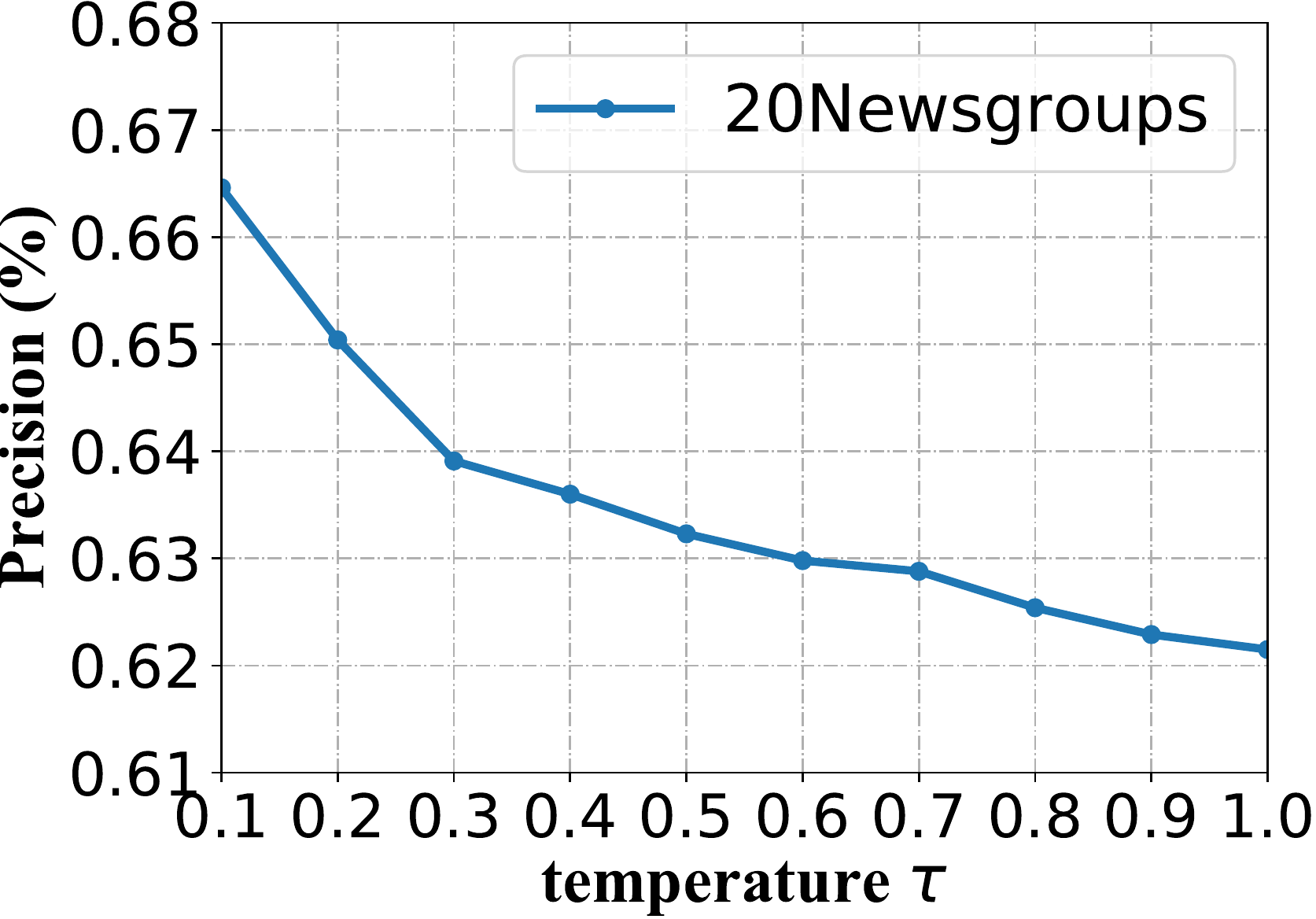}
		\end{minipage}
	}
	\caption{The precision of 64-bit hash codes on three datasets with varying temperature $\tau$ and KL weight $\beta$.}
	\vspace{-6mm}
	\label{fig:parameter_sensitive}
\end{figure}

\section{Additional Experiments}

\paragraph{Comparing with RBSH and PairRec} As mentioned before, the reason we do not directly compare our method with RBSH \cite{hansen2019unsupervised} and PairRec \cite{hansen2020unsupervised} is that their data processing methods are different from the mainstream methods ({\it e.g.}, VDSH, NASH, GMSH, NbrReg, AMMI and CorrSH). To further compare our method with them, we evaluate our model on three datasets that are published by RBSH\footnote{https://github.com/casperhansen/RBSH}. The results are illustrated in Table \ref{table:compare_RBSH}. We observe that our method achieves the best performances in most experimental settings, which further confirms the superiority of simultaneously preserving the semantics and similarity information in a more principled framework.

% \paragraph{Analysis of Semantic Information}
% To illustrate the semantic information learned from the word embedding matrix $E \in \mathbb{R}^{d \times |V|}$ in decoder network, we select the five nearest words based on the cosine similarity of their compact representation, with the result shown in Table \ref{table:semantic_information}. It can be seen that, compared with NASH, our model effectively groups semantically-similar words together in the learned vector space, demonstrating the preferable ability to capture semantic information of documents.

\paragraph{Parameter Sensitivity} To understand the robustness of our model, we conduct a parameter Sensitivity analysis of $\tau$ and $\beta$ in Figure \ref{fig:parameter_sensitive}. Compared with $\beta = 0$ (without using neighborhood information), models with $\beta \neq 0$ improve performance significantly, but gradually performs steadily as $\beta$ getting larger, which once again confirms the importance of simultaneously modeling semantic and neighborhood information. As for temperature coefficient $\tau$ used in variational encoder, our model performs steadily with various values of $\tau$ in the Reuters dataset. But in TMC and 20Newsgroups, increasing $\tau$ would deteriorate the model performance. Generally speaking, the model can achieve better performance with smaller $\tau$ ({\it i.e.}, steeper sigmoid function). As we utilize $0.5$ as the threshold value, steeper sigmoid functions make it easier to distinguish hash codes.

\end{document}